%% file: 6652.tex
\documentclass{aa} 
\usepackage{lscape}
\usepackage{graphicx}
\usepackage{txfonts}
\begin{document}
\title{The brightest stars of the $\sigma$ Orionis cluster}
%
%
\author{J. A. Caballero\inst{1,2,3}\fnmsep\thanks{Alexander von Humboldt Fellow
at the Max-Planck-Institut f\"ur Astronomie, \email{caballero@mpia.de}.}} 
%
%
\institute{
Max-Planck-Institut f\"ur Astronomie, K\"onigstuhl 17, D-69117 Heidelberg,
Germany 
\and 
Isaac Newton Group, Apartado 321, E-38700 Santa Cruz de La Palma, Canary
Islands, Spain
\and 
Instituto de Astrof\'{\i}sica de Canarias, E-38205 La Laguna, Tenerife, Spain
}
\date{Received 27 October 2006; accepted 2 January 2007}

\abstract
{The very young $\sigma$ Orionis cluster ($\sim$ 3\,Ma) is a cornerstone for
the understanding of the formation of stars and substellar objects down to
planetary masses.  
However, its stellar population is far to be completely known.}   
{This study has the purpose of identyfing and characterising the most massive
stars of $\sigma$ Orionis to complement current and future deep searches for
brown dwarfs and planetary-mass objects in the cluster.}      
{I have cross-correlated the sources in the Tycho and 2MASS catalogues in a
region of 30\,arcmin radius with centre in the O-type star $\sigma$ Ori A. 
In the area, I have studied the membership in the Ori OB 1b Association
of the brightest stars in the optical using astrometric, X-ray and infrared and
optical photometric data from public catalogues and spectroscopic data from the
literature.}    
{A list of 26 young stars, four candidate young stars and 16 probable foreground
stars has arised from the study.
Seven young stars probably harbour discs (four are new).
There is no mass-dependence of the disc frequency in the cluster.
I have derived for the first time the mass spectrum in $\sigma$ Orionis from 1.1
to 24\,M$_\odot$ ($\alpha$ = +2.0$^{+0.2}_{-0.1}$; roughly Salpeter-like). 
I have also provided additional proofs on the existence of several spatially
superimposed stellar populations in the direction of $\sigma$ Orionis.
Finally, the cluster may be closer and older than previously
considered\footnote{Tables \ref{estrellas_uno} and \ref{estrellas_dos}  
are only available in electronic form at the CDS via anonymous ftp to
cdsarc.u-strasbg.fr (130.79.128.5) or via {\tt
http://cdsweb.u-strasbg.fr/cgi-bin/qcat?J/A+A/}.}.} 
{}
\maketitle
%

\section{Introduction}

The constellation of Orion, the Hunter, is dominated by the stellar components
of the Ori OB 1 complex.  
The complex is probably related to the Barnard Loop, a huge feature with
H$\alpha$ in emission that extends up to Eridanus, and to a hole in the
distribution of gas H {\sc i}, which may have been originated from several
supernova events that happened about 4\,Ma ago (Brown, de Geus \& de
Zeeuw 1994).  
The very young Orion complex is a set of four subgroups or associations of very
early type stars, named Ass Ori OB 1a, 1b, 1c and 1d.  
Each of them possesses slightly different locations, ages (1--10\,Ma) and
heliocentric distances (300--500\,pc).  
Blaauw (1964), Warren \& Hesser (1978), Goudis (1982), de Geus et al. (1990),
Brown et al. (1994) and other authors have described the boundaries
between subgroups and their characteristics.  

The $\sigma$ Orionis cluster lies within the Ori OB 1b Association, also
known as the Orion Belt. 
This association contains three famous perfectly aligned very bright
stars: \object{Alnitak} ($\zeta$ Ori), \object{Alnilam} ($\epsilon$ Ori) and
\object{Mintaka} ($\delta$ Ori).  
It has been suggested that the age, distance and radial velocity of the stellar
components vary across the association (Hardie et al. 1964; Warren \& Hesser
1977a,b; Guetter 1981; Gieseking 1983; Genzel \& Stutzki 1989). 
The east part of the Ori OB 1b Association, with Alnitak, the \object{Horsehead
Nebula}, the \object{Flame Nebula} (the Orion B cloud, associated to \object{NGC
2024}) and the H {\sc ii} region \object{IC 434}, would be the farthest and
youngest one. 
Actually, the star formation is still underway close to the Horsehead Nebula,
which mane is illuminated by the $\sigma$ Ori star\footnote{Through this paper
and for clarity, I am using the full and shortened names for stars and clusters,
respectively; i.e. the $\sigma$ Ori star system, the $\sigma$ Orionis cluster.},
the fourth brightest star in the Orion Belt. 

Garrison (1967) and Lyng\aa~(1981) were the first investigators to recognize the
$\sigma$ Orionis cluster. 
It went unnoticed until Wolk (1996) and Walter, Wolk \& Sherry (1998) detected
and over-density of X-ray sources and a population of pre-main-sequence
low-mass stars surrounding the centre of the cluster (see also: Walter et al.
1997; Wolk \& Walter 2000). 
Although it is not as extremely young as the Trapezium cluster in the Ori OB 1d
Association (the Orion Sword) and is not as nearby to the Sun as the rich
star-forming regions in the Southern Hemisphere (i.e. Chamaeleon, Ophiuchus),
the $\sigma$ Orionis cluster has the advantage of possessing simultaneously a
moderate youth, a relative short heliocentric distance and, especially, a very
low interstellar extinction (Lee et al. 1968; B\'ejar et al. 2004b). 
This fact may be due to strong winds, ultraviolet radiation and/or turbulence
generated by the O-type star in the cluster centre, which would sweep away the
gas and dust. 
Furthermore, several authors have suggested that $\sigma$ Orionis is in reality
an H {\sc ii} region (e.g. Reipurth et al. 1998). 

There are different determinations of the age of the $\sigma$ Orionis cluster,
based on different methodologies: ($i$) from the comparison between
mid-resolution real and theoretical spectra surrounding the Li {\sc i}
$\lambda$6707.8\,\AA~youth-indicator line, ($ii$) from the presence of very
early-type stars, ($iii$) from the disc frequency around stars and ($iv$)
from fits to theoretical isochrones in colour-magnitude diagrams. 
All the determinations are directed towards a narrow age interval, between 0.5
and 8\,Ma. 
In Table \ref{jardina_age}, I summarize several age determinations found in the
literature. 
The most probable age of the cluster is about 3\,Ma, with conservative --2 and
+5\,Ma error-bars. 

   \begin{table}
      \caption[]{Different determinations of the age of the $\sigma$ Orionis cluster.}
         \label{jardina_age}
     $$ 
         \begin{tabular}{l r}
            \hline
            \hline
            \noalign{\smallskip}
Age (Ma)		& Reference 	\\
            \noalign{\smallskip}
            \hline
            \noalign{\smallskip}
8			& Blaauw (1964) \\
$<$5.1$^a$		& Warren \& Hesser (1978) \\
7			& Blaauw (1991) \\
1.7$\pm$1.1		& Brown et al. (1994) \\
4.2 $^{+2.7}_{-1.5}$	& Oliveira et al. (2002) \\
2--4$^b$		& Zapatero Osorio et al. (2002b) \\
2.5$\pm$0.3		& Sherry, Walter \& Wolk (2004) \\
3.5$\pm$3.0		& Hern\'andez et al. (2005) \\
4--6			& Brice\~no et al. (2005) \\
6.6			& Kharchenko et al. (2005) \\            
$\lesssim$4--6$^c$	& Sacco et al. (2006) \\            
	\noalign{\smallskip}
            \hline
         \end{tabular}
     $$ 
\begin{list}{}{}
\item[$^{a}$] Age not rotation-corrected (upper limit: 5.1\,Ma).
\item[$^{b}$] Upper limit: 8\,Ma; lower limit: 1\,Ma. 
\item[$^{c}$] They also found three low-mass high probability members that
are older than $\sim$10\,Ma.
\end{list}
   \end{table}

The heliocentric distance to $\sigma$ Ori from the trigonometric parallax
measured by the Hipparcos mission is 352$^{+166}_{-168}$\,pc.  
There are also different determinations of the heliocentric distance to the Ori
OB 1b Association and to several $\sigma$ Orionis members (Brown et al. 1994,
1999; Perryman et al. 1997; de Zeeuw et al 1999; Brown et al. en prep.). 
Published distances vary between 360$^{+70}_{-60}$\,pc and 473$\pm$33\,pc.
The most recent determinations of the heliocentric distance to the $\sigma$
Orionis cluster were shown by Hern\'andez et al. (2005), who determined the
distances from parallax, 443$\pm$16\,pc, and from colour-magnitude diagrams,
392$\pm$20\,pc, and by Sherry, Walter \& Wolk (2004), who determined an
heliocentric distance of 440$\pm$40\,pc. 
I will use the value measured by Brown et al. (1994) of 360$^{+70}_{-60}$\,pc. 

B\'ejar et al. (1999) opened the era the deep photometric searches in the
cluster and detected the first substellar objects in $\sigma$ Orionis (they have
not stable hydrogen burning in their cores, in contrast to stars). 
Posterior studies have gone deeper, discovering a wealthy population of brown
dwarfs (B\'ejar et al. 2001, 2004b; Scholz \& Eisl\"offel 2004; Caballero
et al. 2004; Kenyon et al. 2005) and even objects with masses in the planetary
domain (below the deuterium burning mass limit) (Zapatero Osorio et al. 2000;
Gonz\'alez-Garc\'{\i}a et al. 2006; Caballero et al. 2007a).
The $\sigma$ Orionis cluster is to date the star-forming region with the largest
amount of candidate planetary-mass objects (27).
Out of them, 12 have low-resolution spectroscopy (Mart\'{\i}n et al. 2001;
Barrado y Navascu\'es et al. 2001) and three harbour surrounding discs
(Caballero et al. 2007b).
The cluster has now turned out to be not only the region with the largest
stellar density in the Ori OB 1b Association, but one of the most interesting
regions in the Milky Way to study and understand the formation of stars, brown
dwarfs and, especially, isolated planetary-mass objects. 

In spite of the importance of the cluster, its stellar population is
paradoxically poorly known.
General studies in the Ori OB 1b Association (e.g Warren \& Hesser 1977a, 1977b,
1978), very wide searches with prism-objective and Schmidt plates (Haro \&
Moreno 1953; Kogure et al. 1989; Wiramihardja et al. 1989, 1991; Nakano et al.
1995) and the {\em ROSAT}-based search by Wolk (1996) provided most of the known
cluster members at the end of last century.
However, these surveys were obviously biased towards the detection of active
Ori OB1 1b stellar members.
Some efforts have been recently carried out to improve our knowledge of the
stellar population in $\sigma$ Orionis.
Nonetheless, the bias towards H$\alpha$ emitters (Weaver \& Babcock 2004),
large-amplitude photometric variables (Scholz \& Eisl\"offel 2004) or low-mass
objects (Tej et al. 2002; Sherry et al. 2004) is still prevalent.
The summit of our relative poor knowledge of the stellar and high-mass
brown-dwarf population in the cluster is maybe the case of \object{SE 70}.
Scholz \& Eisl\"offel (2004) detected it at 4.6\,arcsec to a previously known
$\sim$ 5\,$M_{\rm Jup}$-mass candidate planetary-mass object, \object{S\,Ori 68}
(Zapatero Osorio et al. 2000).
Caballero et al. (2006a) have recently confirmed the membership in the cluster
of \object{SE 70}, which is an X-ray-emitter brown dwarf\footnote{Burningham
et al. (2005) also measured the equivalent width of the Na {\sc i}
$\lambda\lambda$8183,8195\,\AA~doublet and the radial velocity of SE 70,
which are very similar to those of other confirmed young brown dwarfs in the
$\sigma$ Orionis cluster.} and may form together with S\,Ori 68 the widest
planetary system ever found (1700$\pm$300\,AU). 
The recent discovery of at least two spatially coincident stellar
populations in the area makes current and future studies more complex.
The superimposed stellar components, firstly showed by Jeffries et al. (2006)
and confirmed by Sacco et al. (2006), are kinematically separated in radial
velocity and have different mean ages.
Furthermore, $\sigma$ Orionis itself might not be a member of any of the known
Ori OB1 subgroups, but be part of a new one.
For example, the simple fact of the Horsehead Nebula in the Ori OB 1b
Association being a back-illuminated absorption nebula may suggest that it is at
a different heliocentric distance than $\sigma$ Orionis.

A complete study of the stellar population in the $\sigma$ Orionis cluster is
absolutely needed to complement current and future deep searches of brown dwarfs
and planetary-mass objects in the area (e.g. to compare spatial 
distributions, disc frequencies or slopes of the mass spectrum at different
mass intervals and separations to nearby massive stars).
In this paper, I will describe an optical-near infrared search for the
most massive stars in the area of $\sigma$ Orionis, supported by
astrometric, X-ray and mid-infrared measurements. 
The search and characterization of intermediate- and low-mass stars in the
cluster will be presented in a forthcoming paper.

\section{Analysis}

A cross match between the Tycho-2 and 2MASS catalogues will lead to the
identification of the brightest stars (in the optical) of the Ori OB 1b
Association in the direction of the $\sigma$ Orionis cluster. 
The collection of additional data from different archives
related to mid-infrared and X-ray space missions and from the literature will 
provide information on the nature and age of the stars.

\subsection{The Tycho-2/2Mass correlation}

\begin{figure}
\centering
\includegraphics[width=0.49\textwidth]{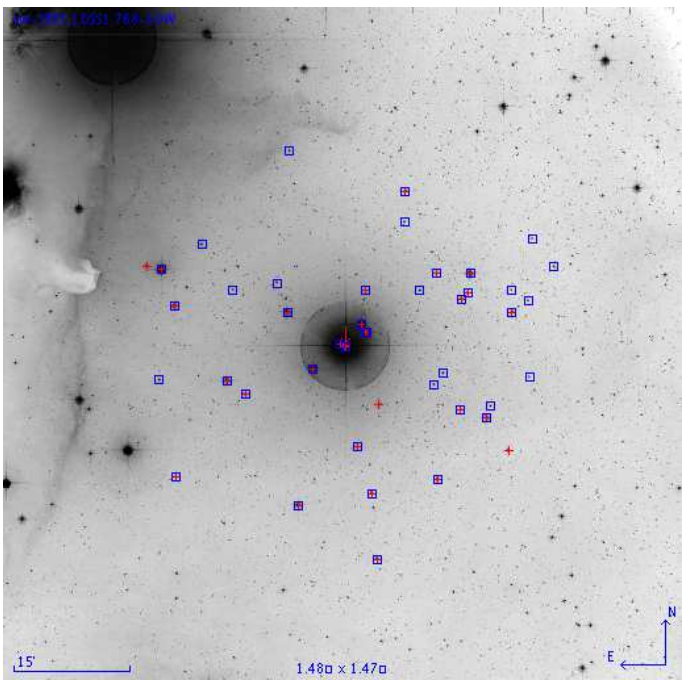}
\caption{Inverse-colour DSS-1 blue-band (photographic $B_J$) image centred in
$\sigma$ Ori A (marked with an arrow). 
Size is 1.5 $\times$ 1.5\,deg$^2$.
North is up and East is left.
Sources in the Tycho-1 and Tycho-2 are marked with (red) crosses and (blue)
squares, respectively (in colour in the electronic version). 
East and north-east of the field of view are dominated by the star-forming
region complex formed by Alnitak, the Flame Nebula
(NGC 2024), the Horsehead Nebula and the IC 434 H {\sc ii}
region. 
}
\label{F01}
\end{figure}

For the present work, I have used the web-based sky atlas Aladin (v3.600;
Bonnarel et al. 2000; Fernique et al. 2004).
I have loaded all the sources catalogued in the Tycho-2 and the
2MASS Point Source catalogues (H{\o}g et al. 2000 and Cutri et al. 2003,
respectively) in a circle of 30\,arcmin radius centred in $\sigma$ Orionis
A+B+IRS1 (the close triple OB system formed by 
\object{$\sigma$ Ori A}, \object{$\sigma$ Ori B} and \object{$\sigma$ Ori
IRS1} that gives the name to the cluster). 
In the studied area of $\sim$ 2\,830\,arcmin$^2$, there are 41 Tycho-2 and about
50\,000 2MASS catalogued sources. 
Previous studies in the region pointed out that 30\,arcmin ($\sim$ 3\,pc at the
distance of $\sigma$ Orionis) is a rough but valid approximation to the cluster
radius (Sherry et al. 2004; B\'ejar et al. 2004a). 
Many bright stars out of this radius may be related to the youngest
star-forming regions to the east of the Ori OB 1b Association or to the
Alnilam cluster to the west (see e.g. Wolk 1996).  
The B1.5V-type star \object{HD 37744} and the Horsehead Nebula are in the
vicinity of the studied region. 

The positional cross match in the area with the Aladin Cross match tool between
the Tycho-2 and 2MASS catalogues provided a list of 41 stellar sources. 
Therefore, the 2MASS near-infrared counterparts of all the Tycho-2 sources in
the area were identified.  
I used the default cross-match threshold of 4\,arcsec.
Except in one case, the agreement between the Tycho-2 and 2MASS coordinates was
better than 0.5\,arcsec (and better than 0.2\,arcsec in 78\,\% of the cases). 
The difference between the Tycho-2 and 2MASS coordinates of the outlier was
2.4\,arcsec. 
This source (TYC 4770 1432 1), with a magnitud in the reddest
Tycho band $V_T$ = 12.3$\pm$0.2\,mag, is one of the faintest stars in the
sample. 
The deviation may be ascribed to the low signal-to-noise ratio in the Tycho
catalogue, which 90\,\%-completeness magnitude in the $V$ band is about
11.5\,mag (H{\o}g et al. 2000).

Figure \ref{F01} shows a $\sim$ 1.5 $\times$ 1.5\,deg$^2$ blue digitized
photographic plate, covering the region of 30\,arcmin centred in $\sigma$ Ori A
and adjacent areas.  
The 41 correlated stars are marked in the view window with squares.
For each star, it is available photometry in five bands ($B_T V_T J H K_{\rm
s}$) and accurate proper motion measurements. 
In the upper parts of Tables \ref{estrellas_uno} and \ref{estrellas_dos}, I
provide their names and the photometric and astrometric data,
respectively\footnote{Tables \ref{estrellas_uno} and \ref{estrellas_dos}  
are only available in electronic form at the CDS via anonymous ftp to
cdsarc.u-strasbg.fr (130.79.128.5) or via {\tt
http://cdsweb.u-strasbg.fr/cgi-bin/qcat?J/A+A/}.}.   
The stars in the Tycho-1 catalogue (ESA 1997), superseded in most applications
by Tycho-2, are marked with crosses in Figure \ref{F01}. 
There are four additional stars that are in Tycho-1 and not in Tycho-2.

\subsection{Additional data}

\subsubsection{{\em IRAS} space telescope}
\label{iras}

   \begin{table}
      \caption[]{Mid-infrared sources among the stars in the Tycho-2/2Mass
      correlation.} 
         \label{estrellas_mIR}
     $$ 
         \begin{tabular}{lcccc}
            \hline
            \hline
            \noalign{\smallskip}
Name 			& $F_{12}$  	& $F_{25}$  	& $F_{60}$  	& $F_{100}$ \\  
 			& [Jy]		& [Jy]		& [Jy]		& [Jy]	\\  
            \noalign{\smallskip}
            \hline
            \noalign{\smallskip}
$\sigma$ Ori A+B+IRS1$^{a}$& 4.5$\pm$0.2& 15$\pm$2 	& 15$\pm$2 	& 15$\pm$2 \\ 
HD 37699 		& 0.29:		& 0.45$\pm$0.09 & 1.5$\pm$0.2 	& 4.9: 	\\ 
HD 294268 		& 0.87$\pm$0.11 & 2.74$\pm$0.16 & 1.50$\pm$0.10 & 22: 	\\ 
            \noalign{\smallskip}
            \hline
         \end{tabular}
     $$ 
\begin{list}{}{}
\item[$^{a}$] Re-measured IRAS photometry (Oliveira \& van Loon 2004).  
\end{list}
   \end{table}

Three out of the 41 correlated stars possess an unambiguous source of the {\em
IRAS} catalogue of Point Sources Version 2.0 (IPAC 1986) at less than
40\,arcsec.  
Their names and average non-colour corrected flux densities, $F_\nu$, for
12, 25, 60 and 100\,$\mu$m are given in Table \ref{estrellas_mIR}. 
The three stars have been investigated as mid-infrared sources with probable
discs 
($\sigma$ Ori A+B+IRS1 -- van Loon \& Oliveira 2004; 
HD 294268 -- Garc\'{\i}a-Lario et al.1990, 1997 and Torres et al. 1995;
HD 37699 -- Oganesyan et al. 1995).

In the studied area there are only four additional {\em IRAS} unambiguous
sources associated to fainter stars and their surrounding envelopes. 
They are \object{TX Ori}, \object{V510 Ori}, \object{IRAS 05358--0238} and
\object{IRAS 05352--0227}.
None of them are in the Tycho-2 catalogue.
However, the correlated star HD 294269 has an unreliable detection at only one
passband, 60\,$\mu$m, and will not be considered.  


\subsubsection{{\em ROSAT}, XMM-{\em Newton} and {\em Chandra} space telescopes}

   \begin{table}
      \caption[]{X-ray sources among the stars in the Tycho-2/2Mass correlation$^a$.}
         \label{estrellas_X}
     $$ 
         \begin{tabular}{lcccccccr}
            \hline
            \hline
            \noalign{\smallskip}
Name 			& Nr.	& WGACAT		& XMM	\\  
 			& 1RXH	&[10$^{-3}$\,s$^{-1}$]&[10$^{-3}$\,s$^{-1}$]\\  
            \noalign{\smallskip}
            \hline
            \noalign{\smallskip}
$\sigma$ Ori A+B+IRS1 	& 129	& 330$\pm$80		& 440$\pm$2		\\ 
$\sigma$ Ori E 		& 77	& --			& 199.4$\pm$1.7		\\ 
HD 37699 		& 4	& 90$\pm$20		& $\times$		\\ 
HD 37525 AB 		& 129   & 19.6$\pm$1.3		& --			\\ 
HD 294272 A+B		& 3     & 3.6$\pm$0.6		& 5.9$\pm$0.4		\\ 
HD 37564 		& 4	& 24$\pm$5		& 9.9$\pm$0.5		\\ 
V1147 Ori 		&$\times$& 70$\pm$40		& $\times$		\\ 
HD 294268 		&$\times$& 3.50$\pm$0.11	& $\times$		\\ 
RX J0539.6--0242 AB	& 142	& 60$\pm$20		& 63$\pm$2		\\ 
HD 294298 		&$\times$& 60$\pm$8		& $\times$		\\ 
4771--0950 		& 1	& --		        & 0.67$\pm$0.18		\\ 
2E 0535.4--0241 	& 160	& 53$\pm$17		& 189$\pm$4		\\ 
            \noalign{\smallskip}
            \hline
         \end{tabular}
     $$ 
\begin{list}{}{}
\item[$^{a}$] Name of optical counterpart, number of events in 1RXH-{\em ROSAT},
count rate (s$^{-1}$) in WGACAT-{\em ROSAT} and count rate (s$^{-1}$) in
the XMM-Newton observations by Franciosini et al. (2006). 
The {\em ROSAT} and XMM-{\em Newton} missions are not sensitive to the same
wavelength interval. 
The symbol ``$\times$'' denotes that the target is not in the area studied by
the X-ray mission or catalogue. 
\end{list}
   \end{table}

At least 13 stars of the sample are X-ray emitters.
They were found in the following catalogues: {\em ROSAT} Source Catalog of
Pointed Observations with the High Resolution Imager, 1RXH ({\em ROSAT} 2000a);
WGACAT version of the ROSAT PSPC Catalogue, WGACAT (White et al. 2000); Second
{\em ROSAT} Source Catalog of Pointed Observations, 2RXP ({\em ROSAT} 2000b).  

All the X-ray emitters, except $\sigma$ Orionis E and the faint star 4771--0950
(Wolk 1996), are present in the WGACAT-{\em ROSAT }catalogue. 
Eight were previously detected in the spatial and spectral analysis of a full
{\sc EPIC} field in the XMM-{\em Newton} observations of the $\sigma$ Orionis
cluster by Franciosini et al. (2006). 
They reported deviations between the coordinates of the optical and the
X-ray sources of less than 1\,arcsec for all the bright stars in their sample.
Additionally, HD 37699 and HD 294268, to the northeast of the area, were also
detected in the X-ray observations of the Orion OB 1b Association with the {\em
ASCA} mission by Nakano et al. (1999).
Voges et al. (1999) considered GSC 04771--00962 and GSC 04771--00961 as possible
HST Guide Star Catalogue optical counterparts of the X-ray {\em ROSAT} source
1RXS J053756.8--023857, which is in reallity adscribed to the bright young star
2E 0535.4--0241.  
Table \ref{estrellas_X} shows the X-ray emitters in the sample, the count rates
and the number of events measured by different instruments and space missions. 
 
The high background produced by the extreme X-ray emitter $\sigma$ Ori
A+B+IRS1 did not allow to recognise $\sigma$ Ori E as an individual WGACAT
source.
As well, the small angular separation between HD 294272 A and B ($\sim$
8\,arcsec) and their similarity in brightness did not permit to discriminate
which is the real X-ray emitter from {\em ROSAT} and XMM-{\em Newton} data.  
However, both $\sigma$ Ori A+B+IRS1 and $\sigma$ Ori E and HD 294272 A and B
systems are resolved in the high-spatial-resolution X-ray images from the {\em
Chandra} Space Telescope archive obtained by Adams et al. (2004) with {HRI-C}. 

The null detection at high energies of the rest of the targets, except in the
case of StHA 50, may be due to the large separation to $\sigma$ Ori A, where
X-ray surveys are usually centred.

\subsubsection{Optical spectroscopy}
\label{optical}

From my searches in the literature, 28 stars have a spectral type determination.
Their spectral types are provided in the last column of Table
\ref{estrellas_dos}. 
Out of them, 13 stars (14 if the Be-type star StHa 50 is also included)
have OBA spectral types. 
Other five stars have F types, and the rest are G- and K-type stars.
The spectral type determinations have been borrowed from the following works:
Schild \& Chaffee (1971), Warren \& Hesser (1978), Guetter (1981), Downes \&
Keyes (1988), Gray \& Corbally (1993), Nesterov et al. (1995), Wolk (1996),
Catalano \& Renson (1998), Alcal\'a et al. (2000), Gregorio-Hetem \& Hetem
(2000) and Caballero (2006). 

There is information about the lithium abundance for seven stars.
In six cases, the equivalent width of the Li {\sc i}
$\lambda$6707.8\AA~suggests that the lithium abundance is primordial
($\log{\epsilon({\rm Li})} \sim$ 3.0), while there is an upper limit of 1.0\,dex
for the $\log{\epsilon({\rm Li})}$ of the G0-type star HD 294269 (Cunha, Smith
\& Lambert 1995; Torres et al. 1995; Alcal\'a et al. 1996, 2000; Caballero
2006).  
See in Cunha et al. (1995) a careful study of the lithium abundance of late
F and early G stars in Orion and how very young ages can be derived for stars
with effective temperatures between $\sim$5200 and $\sim$6400\,K.
 
There are three stars with H$\alpha$ in emission: RX J0539.6--0242 AB
(EW(H$\alpha$) = --2.38\,\AA; Alcal\'a et al. 1996), StHA 50 (Stephenson 
1986; Downes \& Keyes 1988) and HD 294268 (EW(H$\alpha$) = --2\,\AA; Torres et
al. 1995).
As well, seven stars display spectroscopic peculiarities: V1147 Ori
(abnormally strong silicon and europium lines; Catalano \& Renson 1998), HD
37333 (mild Si {\sc ii}; Gray \& Corbally 1993), HD 37699, HD 37686 and HD
294301 (broad --``nebulous''-- absorption due to spinning; Neubauer 1943, Warren
\& Hesser 1978 and Gray \& Corbally 1993, respectively), HD 37525 AB (helium
weak; Nissen 1976) and $\sigma$ Ori E (helium rich and magnetic strong;
Greenstein \& Wallerstein 1958).  
The stars HD 294298 and RX J0539.6--0242 AB have also large values of $v
\sin{i}$ if compared with field solar-like stars (55 and
150\,km\,s$^{-1}$, respectively; Cunha et al. 1995 and Alcal\'a et al. 1996).
Finally, Hunger et al. (1989) measured a $v \sin{i}$ of 160$\pm$20\,km\,s$^{-1}$
in $\sigma$ Ori E.

\subsection{Target classification}

\subsubsection{Young stars}

   \begin{table}
      \caption[]{Very young stars in the Tycho-2/2MASS correlation.}
         \label{estrellas_jovenes}
     $$ 
         \begin{tabular}{lcccccc}
            \hline
            \hline
            \noalign{\smallskip}
Name 			& OB	& X-ray	& Li {\sc i}	& H$\alpha$	& mIR	& var. \\  
            \noalign{\smallskip}
            \hline
            \noalign{\smallskip}
$\sigma$ Ori A+B+IRS1 	& Y	& Y	&		&		& Y	&	\\
$\sigma$ Ori E 		& Y	& Y	&		&		&	& Y	\\
HD 37699     		& Y	& Y	&		&		& Y	&	\\
HD 294271    		& Y	&	&		&		&	&	\\
HD 37525 AB  		& Y	& Y	&		&		&	&	\\
HD 294272 A  		& Y	& Y	&		&		&	&	\\
HD 294272 B  		& Y	& Y	&		&		&	&	\\
HD 37333     		& Y	&	&		&		&	&	\\
HD 37564     		& 	& Y	&		&		&	&	\\
V1147 Ori    		& Y	& Y	&		&		&	& Y	\\
HD 37686     		& Y	&	&		&		&	&	\\
HD 37545     		& Y	&	&		&		&	&	\\
HD 294275    		& Y	&	&		&		&	&	\\
HD 294268		& 	& Y	& Y		& Y		& Y	& Y?	\\
RX J0539.6--0242 AB	& 	& Y	& Y		& Y		&	&	\\
HD 294279    		& 	&	& Y		&		&	&	\\
StHA 50  		& Y?	&	&		& Y		&	&	\\
HD 294298    		& 	& Y	& Y		& 		&	&	\\
4771--0950		& 	& Y	& Y		&		&	&	\\
2E 0535.4--0241		& 	& Y	&		&		&	&	\\
SO120532	      	& 	&	& Y		&		&	&	\\
            \noalign{\smallskip}
            \hline
         \end{tabular}
     $$ 
   \end{table}

The primordial abundance of lithium (``Li {\sc i}''), the line of H$\alpha$ in
appreciable emission in solar-type stars or earlier (``H$\alpha$''), the
mid-infrared flux excess in the {\em IRAS} pass-bands due to a disc (``mIR''),
the very early spectral types (O, B and early A; ``OB''), the X-ray emission
(strong if a solar-type star; ``X''), the high-amplitude photometric variability
(``var.'') and the fast rotation are some known youth features (age
$\lesssim$ 10\,Ma).  
These signatures are common in T Tauri stars and their analogous at intermediate
masses, the Herbig Ae/Be stars (with masses in the interval 1.5\,M$_\odot$ $<$ M
$<$ 10\,M$_\odot$ and spectral types earlier than F; Hillenbrand et al. 1992). 
Out of the 41 stars of the Tycho-2 sample, 21 have at least one youth feature. 
In Table \ref{estrellas_jovenes}, I provide the list of the very young stars.
I will consider that they are bona-fide members of the Ori OB 1b Association, to
which the $\sigma$ Orionis cluster belongs.
Some of the stars display several youth features at a same time, as HD 294268,
which is a mid-infrared source with moderate X-ray and H$\alpha$ emissions and
lithium in absorption.
It is also a probable photometric variable star. 
Warren \& Hesser (1978) suggested the membership to the Ori OB 1b Association
of 14 stars among the subsample of young stars in Table \ref{estrellas_jovenes}.

\subsubsection{Stars with large tangential velocity}
\label{propermotion}

\begin{figure}
\centering
\includegraphics[width=0.49\textwidth]{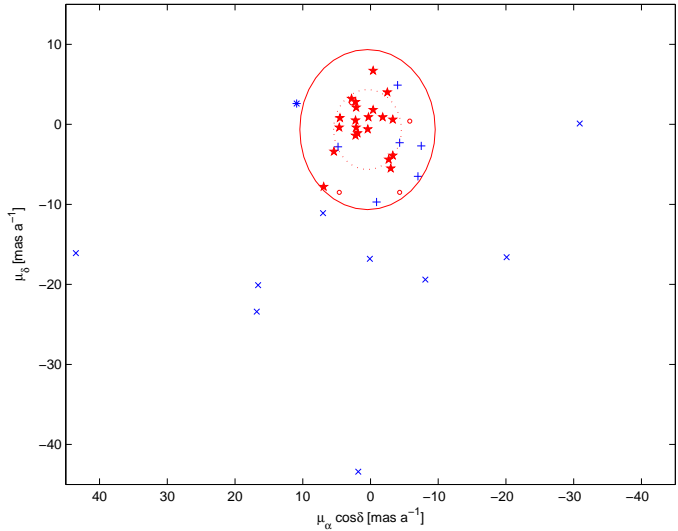}
\caption{Proper motions of the 41 Tycho-2/2MASS-correlation stars. 
Code: 
(red) filled stars, ``$\star$'', young stars; 
(blue) crosses, ``$\times$'', field stars with proper motion different from the
mean proper motion of the Ori OB 1b Association; 
(blue) pluses, ``$+$'', probable K- or M-type foreground stars; 
(red) open circles, ``$\circ$'', possible new Ori OB 1b Association members. 
The dotted and solid big circles (ellipses with the axes scale) denote the
Brown et al. (1999) and this paper astrometric selection criteria,
respectively.
The object marked with an asterisk, ``$\ast$'', is HD 294278, which is a
K-type foreground star with a different proper motion (``$\times$'' and ``$+$''
simultaneously).}   
\label{F02}
\end{figure}

The tabulated mean proper motion of all the subgroups of the
Ori OB1 Association is $\mu_\alpha \cos{\delta}$ = +0.44\,mas\,s$^{-1}$,
$\mu_\delta$ = --0.65\,mas\,s$^{-1}$ (Brown et al. 1999).
de Zeeuw et al. (1999) showed that the tangential velocities of 96\,\% of the
Ori OB1 photometric candidates identified by Brown et al. (1994) deviated less
than 5\,mas\,s$^{-1}$ from that mean proper motion.
However, in the work presented here, there is a large fraction of young
stars that display features of extreme youth and their tangential velocity do
not satisfy the de Zeeuw et al. (1999) proper-motion criterion.
One extreme case is that of the F7-type X-ray-emitter star 4771--0950, with
pEW(Li {\sc i}) = +0.07$\pm$0.01\,\AA~(Caballero 2006), which proper motion
deviates in almost 10\,mas\,s$^{-1}$ from the mean proper motion of
the Ori OB1 Association. 
I have used the value of 10\,mas\,s$^{-1}$ to separate young stars and candidate
Association members from probable foreground stars with larger tangential
velocities. 
This wider ``error circle'' is not caused by larger errors in the proper
motions of the faintest members, but due to a genuine tangential velocity
dispersion (see Section \ref{different}).
Figure \ref{F02} illustrates the astrometric study, showing the selection
criteria by Brown et al. (1999) and the one used here.
The ten stars with tangential velocity different from that of the Ori OB1
Association in more than 10\,mas\,s$^{-1}$ are shown in Table
\ref{estrellas_pm} (they could be at up to 3\,deg from $\sigma$
Ori AB in 1\,Ma).

   \begin{table}
      \caption[]{Stars in the Tycho-2/2MASS correlation with proper motion
      different from that of the Ori OB1 Association in more than
      10\,mas\,s$^{-1}$.} 
         \label{estrellas_pm}
     $$ 
         \begin{tabular}{lccr}
            \hline
            \hline
            \noalign{\smallskip}
Name 			& $\mu_\alpha \cos{\delta} \pm \delta\mu_\alpha \cos{\delta}$ & $\mu_\delta \pm \delta\mu_\delta$ & Sp. \\  
 			& [mas a$^{-1}$] & [mas a$^{-1}$] & type \\  
            \noalign{\smallskip}
            \hline
            \noalign{\smallskip}
HD 294307$^{a}$    	&   +0.1$\pm$1.4 &--16.8$\pm$1.4  & F8 \\
GSC 04771--00621	&   +7.0$\pm$1.6 &--11.1$\pm$1.6  & \\
HD 294278    		&  +10.9$\pm$1.4 &  +2.6$\pm$1.4  & K2 \\
HD 294270$^{b}$    	&  +16.8$\pm$1.9 &--23.4$\pm$1.8  & G0 \\
HD 294269$^{c}$    	&  +43.5$\pm$1.8 &--16.1$\pm$1.8  & G0 \\
HD 294274    		&  --8.1$\pm$3.2 &--19.4$\pm$3.5  & G0 \\
TYC 4771 720 1		& --20.1$\pm$3.9 &--16.6$\pm$4.2  & \\
TYC 4771 661 1		&  +16.6$\pm$4.3 &--20.1$\pm$4.7  & \\
TYC 4770 1432 1		& --30.9$\pm$4.5 &  +0.1$\pm$5.0  & \\
TYC 4770 924 1		&   +1.8$\pm$4.7 &--43.4$\pm$5.1  & \\
            \noalign{\smallskip}
            \hline
         \end{tabular}
     $$ 
\begin{list}{}{}
\item[$^{a}$] Non member of the Ori OB 1b Association in Warren \& Hesser (1978). 
\item[$^{b}$] Uncertain association membership status in Warren \& Hesser
(1978). 
\item[$^{c}$] No detection of Li {\sc i} in absorption and V$_r$ =
+72.8\,km\,s$^{-1}$ in Cunha et al. (1995). 
\end{list}
   \end{table}

\subsubsection{K- and M-type foreground stars}
\label{KandM}

\begin{figure}
\centering
\includegraphics[width=0.49\textwidth]{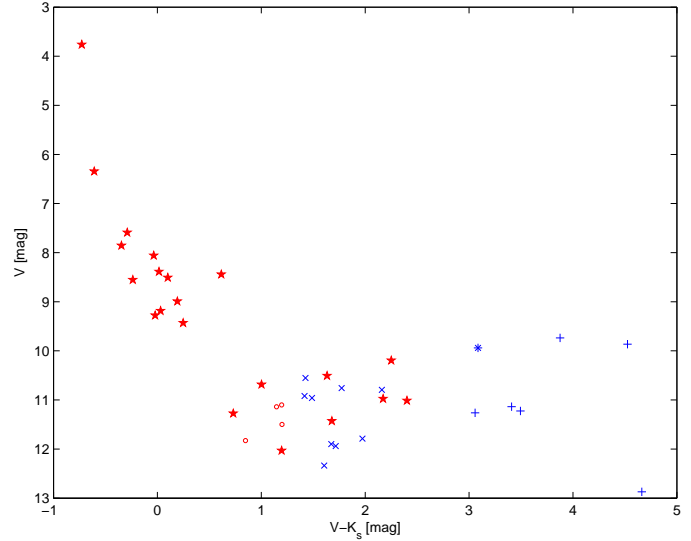}
\caption{Same as Figure \ref{F02} but for the $V_T$ vs. $V_T-K_{\rm s}$ colour-magnitude diagram.}
\label{F03}
\end{figure}

In the $V_T$ vs. $V_T-K_{\rm s}$ colour-magnitude diagram shown in Figure
\ref{F03}, there are two well-differentiated regions separated at $V_T \sim$
9.5\,mag. 
On the one hand, the 13 stars brighter than this value are bluer than
$V_T-K_{\rm s}$ = 1.0\,mag, display youth features and, hence, verify the
astrometric criterion.   
On the other hand, out of the 28 stars fainter than $V_T \sim$ 9.5\,mag, ten are
the proper-motion outliers shown in Table \ref{estrellas_pm}.
There are also seven stars, shown in Table \ref{estrellas_km}, with $V_T-K_{\rm
s}$ colours redder than $\sim$ 3.0\,mag, which are located far from the
photometric cluster sequence defined by the young stars. 
Their $V_T-K_{\rm s}$ colours and $V_T$ apparent magnitudes are typical of 
early/intermediate K dwarfs in foreground. 
Furthermore, Nesterov et al. (1995) determined K spectral type for the three
brightest stars.
Given their brightness (with $J$ magnitudes brighter than 9.5\,mag),
if they were association members, then they should be extraordinary reddened
stars showing the astrophysical properties of Class I objects.
None of them shows X-ray or mid-infrared emissions detectable by present
surveys, nor any spectroscopic youth feature.
On the contrary, the three young members of the Ori OB 1b Association with
$V_T-K_{\rm s}$ colours between 2.0 and 2.5\,mag and of similar brightness in the
$V_T$ band to the stars in Table \ref{estrellas_km} are among the strongest
X-ray emitters in the $\sigma$ Orionis cluster (RX J0539.6--0242 AB, 2E
0535.4--0241 and HD 294298).  

   \begin{table}
      \caption[]{Probable K- and M-type foreground stars in the Tycho-2/2MASS
      correlation.} 
         \label{estrellas_km}
     $$ 
         \begin{tabular}{lr}
            \hline
            \hline
            \noalign{\smallskip}
Name 			& Sp. type \\  
            \noalign{\smallskip}
            \hline
            \noalign{\smallskip}
HD 294277$^{a}$    	& K2	\\
HD 294278$^{b}$    	& K2	\\
HD 294280$^{a}$    	& K5	\\
TYC 4770 1018 1    	& 	\\
TYC 4771 1012 1    	& 	\\
TYC 4771 934 1    	& 	\\
TYC 4771 1468 1    	& 	\\
            \noalign{\smallskip}
            \hline
         \end{tabular}
     $$ 
\begin{list}{}{}
\item[$^{a}$] They were found by the Midcourse Space Experiment satellite at
8.3\,$\mu$m (Kraemer et al. 2003). 
\item[$^{b}$] It does not satisfy the astrometric selection criterion
either. 
\end{list}
   \end{table}

\subsubsection{Possible new Ori OB 1b Association members}

There are four stars that have not been classified as young stars, proper-motion
outliers or foreground K-type stars, and that follow the photometric sequence of
the normal young stars.
They are tabulated in Table \ref{estrellas_nuevas}.
There is available spectroscopic information only for the fast-rotating F2-type
star HD 294301 (Gray \& Corbally 1993).
The star GSC 04771--00962 might be related to an unidentified X-ray event in the
Second {\em ROSAT} PSPC Catalog, at 0.7\,arcmin to the north of the star. 
I will consider the four stars as possible new Ori OB 1b Association members.

   \begin{table}
      \caption[]{Possible new Ori OB 1b association member stars in the
      Tycho-2/2MASS correlation.} 
         \label{estrellas_nuevas}
     $$ 
         \begin{tabular}{lr}
            \hline
            \hline
            \noalign{\smallskip}
Name 			& Sp. type \\  
            \noalign{\smallskip}
            \hline
            \noalign{\smallskip}
HD 294301$^{a}$    	& F2V(n)\\
GSC 04771--00962    	& 	\\
TYC 4770 1261 1    	& 	\\
TYC 4770 1129 1    	& 	\\
            \noalign{\smallskip}
            \hline
         \end{tabular}
     $$ 
\begin{list}{}{}
\item[$^{a}$] Uncertain association membership status in Warren \& Hesser
(1978). 
\end{list}
   \end{table}

\subsection{Binary systems}
\label{sigoriab}

Among the studied targets, there are at least four close multiple systems.
They are $\sigma$ Ori A+B+IRS1, HD 37525 AB, HD 294272 A and B and the
spectroscopic binary RX J0539.6--0242 AB. 
Both Tycho-2 and 2MASS observations were able to resolve the HD 294272 system.
It is not the case, however, of the other three stars.
The secondary at 0.45$\pm$0.4\,arcsec of HD 37525 A is more than 0.5\,mag
fainter in the $H$ band than the primary, so the contribution of the B component
to the total flux at optical wavelengths may not be appreciable.
The same occurs, in an extreme case, with $\sigma$ Ori IRS1, which is more
than 5\,mag fainter than $\sigma$ Ori AB in the $H$ band (Caballero 2005,
2006). 
However, the optical flux of $\sigma$ Ori B is only three times less than that
of $\sigma$ Ori A.
Although the short separation between both components (of 0.250\,arcsec in
J1991.25), it is possible to estimate the optical magnitudes of $\sigma$ Ori A
and B from Hipparcos data. 
The combined $H_P$ magnitude of $\sigma$ Ori AB is 3.681$\pm$0.016\,mag, while
the magnitude difference between them, $\Delta H_p$, is 1.21$\pm$0.05\,mag.
Using the $B-H_P$ and $H_P-V$ colours of blue Hipparcos stars as a reference, I
have estimated that the Johnson $B$- and $V$-band magnitudes of $\sigma$ Ori A
and B are 3.85 and 4.10\,mag, and 5.18 and 5.26\,mag, respectively, with errors
of $\sim$ 0.05\,mag.
I have used the relationships $\Delta m_{A+B} = m_B - m_A$ and $m_A = m_{A+B} +
2.5 \log{\left( 1 + 10^{-\frac{\Delta m_{A+B}}{2.5}} \right)}$.
The $V_A$ and $V_B$ estimations nicely agree with $V$-band adaptive optics
observations of $\sigma$ Ori AB by ten Brummelaar et al. (2000).
The spectroscopic binary RX J0539.6--0242 AB cannot be resolved by imaging with
current technology.
Tovmassian et al. (1991) also proposed a suspected binary among the correlated
stars, HD 37564, but it has not been confirmed.

\subsection{Missing stars}

\subsubsection{Young stars in the Tycho-1 catalogue}
\label{tycho-1}

The Tycho-2 catalogue is not complete.
As mentioned before, there are four stars in Tycho-1 that are not in Tycho-2.
Their basic properties (photometry, astrometry, spectral types) are tabulated in
the bottom parts of Tables \ref{estrellas_uno} and \ref{estrellas_dos}.  
The stars are brighter than or as bright as the Tycho $V_T$ completeness
magnitude. 
Furthermore, one of them is $\sigma$ Ori D, which is one of the
brightest stars in the $\sigma$ Orionis cluster. 
The four stars roughly follow the cluster sequence in the $V_T$ vs. $V_T-K_{\rm
s}$ colour-magnitude diagram. 
Their proper motion measurements are affected by effects external to the
Hipparcos satellite (like the glare of $\sigma$ Ori A+B+IRS1 in the case of
$\sigma$ Ori D or the background emission of the Horsehead
Nebula in the case of HD 294297). 
Errors in their proper motions in Tycho-1 are larger than 5\,mas\,a$^{-1}$ and
up to 30\,mas\,a$^{-1}$, and Tycho-2 not even tabulates them. 
Table \ref{estrellas_dos} shows the tangential velocities of the four stars
provided by H{\o}g et al. (1998), which matches at the level of a few
mas\,a$^{-1}$ with the USNO-B1 and NOMAD1 catalogues (Monet et al. 2003 and
Zacharias et al. 2004, respectively).

   \begin{table}
      \caption[]{Young stars in the Tycho-1 catalogue and not in the Tycho-2
      catalogue.} 
         \label{estrellas_tycho1}
     $$ 
         \begin{tabular}{lcccr}
            \hline
            \hline
            \noalign{\smallskip}
Name 		& OB	& X-ray	& Li {\sc i} 	& Sp. type \\  
            \noalign{\smallskip}
            \hline
            \noalign{\smallskip}
$\sigma$ Ori D$^{a}$& Y & Y	&		& B2V \\ %
HD 294273 	& Y	&	&		& A3 \\ %
HD 294297 	& 	&	& Y		& F6--8 \\ %
            \noalign{\smallskip}
            \hline
         \end{tabular}
     $$ 
\begin{list}{}{}
\item[$^{a}$] $v \sin{i}$ = 180$\pm$20\,km\,s$^{-1}$ (Hunger et al. 1989).
\end{list}
   \end{table}

Out of the four missing stars, three display youth features (Cunha et al. 1995;
Sanz-Forcada et al. 2004; Caballero 2005, 2006).
The young missing stars are shown in Table \ref{estrellas_tycho1}.
Very interestingly, HD 294297 has both a primordial lithium abundance and a
proper motion that clearly deviates from the rest of the Ori OB 1b Association members.
Its tangential velocity is ($\mu_\alpha \cos{\delta}$, $\mu_\delta$) =
(+22.7$\pm$2.6, --20.5$\pm$1.7)\,mas\,a$^{-1}$ from H{\o}g et al. (1998) and
(+22.1$\pm$1.6, --23.9$\pm$1.4)\,mas\,a$^{-1}$ from the NOMAD1 catalogue. 
The remaining missing target, the G0-type star HD 294276, does not display any
youth feature and has the largest tangential velocity among the stars studied in
this work, so it is likely a foreground solar-like star.

\subsubsection{$\sigma$ Ori C}

The Hipparcos mission was mostly sensitive to the blue optical wavelengths.
There are no stars brighter than $J$ = 9.0\,mag in the 2MASS catalogue and in
the surveyed area without optical counterpart in the Tycho catalogues. 
However, there are 12 2MASS stars with magnitudes 9.0 $< J \le$
10.0\,mag not found in Tycho-1 or 2.  
Among them, there are four foreground K- and M-type stars (based on
mid-resolution optical spectroscopy and/or proper motion; Caballero 2006), five
cluster members and three stars of unknown nature.
Out the five young stars, the brightest one is the A2-type star $\sigma$ Ori C,
which is about 2.7\,mag brighter in the $V$ band that the 90\,\%-completeness
magnitude of the Tycho-2 catalogue.
Its absence in the Tycho catalogues may be due to the same reason as $\sigma$
Ori D in the optical and $\sigma$ Ori E in X rays: the glare of $\sigma$ Ori
AB. 
I will ``recover'' $\sigma$ Ori C as one of the brightest $\sigma$ Orionis
cluster members.

\section{Discussion}

To sum up, out of the 41 stars in the Tycho-2 catalogue at angular separations
less than 30\,arcmin of $\sigma$ Ori A, 21 stars have youth features according
to my search in catalogues and in the literature.
I have also computed the $B$- and $V$-band magnitudes of $\sigma$ Ori A and B
separately and considered four additional bright young stars (three from the
Tycho-1 catalogue plus $\sigma$ Ori C) and four new candidate Ori OB 1b
Association members.
This makes a list of 26 bright young stars and four candidate young stars in the
studied area. 
Out of them, nine are in multiple systems: $\sigma$ Ori A, B, C, D and E,
HD 294272 A and B, HD 37525 AB and RX J0539.6--0242 AB.
It indicates that the minimum frequency of multiplicity of the brightest stars
is 35\,\% (the low-mass stars and substellar objects are not taken into
account).
Next, I will characterize the sample of young stars.

\subsection{Individual masses and the $V$ vs. $B-V$ diagram}

For each of the 30 bright young stars and candidates, I have derived their
masses from the Johnson $B$- and $V$-band and theoretical stellar evolution
isochrones from the Geneva database (Lejeune \& Schaerer 2001).
Johnson magnitudes were derived from Tycho $B_T$ and $V_T$ magnitudes using the
relations $V = V_T - 0.090 (B_T - V_T)$ and $B - V = 0.850 (B_T - V_T)$, which
are valid in the colour interval --0.2\,mag $< B_T - V_T <$ 1.8\,mag (ESA 1997).
For $\sigma$ Ori A and B, I have used the $B$ and $V$ magnitudes derived in
Section \ref{sigoriab}, while for $\sigma$ Ori C, I have used the photometry
from Greenstein \& Wallerstein (1958).

\begin{figure}
\centering
\includegraphics[width=0.49\textwidth]{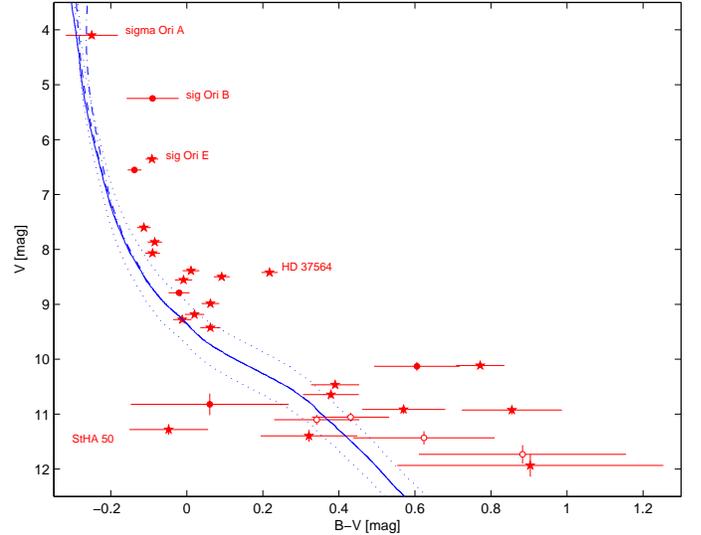}
\caption{$V$ vs. $B-V$ colour-magnitude diagram of the young stars in the
Tycho-2/2MASS correlation (filled stars), the four new probable Ori OB 1b
Association members (open circles), $\sigma$ Ori B, C and the three young stars
in the Tycho-1 recovery (filled circles).
Solid, dashed and dot-dashed lines (in overlapping) denote the 1-, 3- and 10-Ma
isochrones of the Geneva group at the most probable heliocentric cluster
distance, and the dotted lines indicate the 3-Ma isochrones at the minimum
(300\,pc) and maximum (430\,pc) possible cluster distances. 
Several target are labeled.
}
\label{F04}
\end{figure}

The $V$ vs. $B-V$ colour-magnitude diagram depicted in Figure \ref{F04}
illustrates the mass derivation from the models.
In particular, I have used the basic grid for solar metalicity Z = 0.02
(Schaller et al. 1992) and conservative possible intervals in the age
(3$^{+7}_{-2}$\,Ma) and in the heliocentric distance (360$^{+70}_{-60}$\,pc) of
$\sigma$ Orionis. 
The most probable masses for each target and for each age-distance pair was
computed from the mass-$M_V$ relationship given by the models.
The masses with their uncertainties for each target are shown in Table
\ref{estrellas_masas}.
The edge of the conservative error bars correspond to the extreme cases of
a younger closer cluster (1\,Ma, 300\,pc) and for an older farther cluster
(10\,Ma, 430\,pc), which in contrast are close to be ruled out by the isochronal
matches to low-mass stars.
The extended grid with high mass loss for massive stars (Meynet et al. 1994)
provided identical results, except for $\sigma$ Ori A, which derived mass
slightly varied within the uncertainties of the age and heliocentric distance of
the cluster. 
The masses of HD 37525 AB and RX J0539.6--0242 AB were not corrected from
binarity. 

   \begin{table}
      \caption[]{Masses of the young stars.}
         \label{estrellas_masas}
     $$ 
         \begin{tabular}{lcr}
            \hline
            \hline
            \noalign{\smallskip}
Name 			& Mass  		& Remarks$^{a}$ \\  
 			& [M$_\odot$] 		& 		\\  
            \noalign{\smallskip}
            \hline
            \noalign{\smallskip}
$\sigma$ Ori A 		& 18$\pm$6		& XX, (mIR) 	\\ %
$\sigma$ Ori B 		& 12$\pm$3	  	& XX 		\\ %
$\sigma$ Ori E 		&  7.4$^{+1.5}_{-1.4}$	& He rich, XX 	\\ %
$\sigma$ Ori D		&  6.8$^{+1.8}_{-1.2}$	& 		\\ %
HD 37699     		&  4.4$^{+0.8}_{-0.7}$	& mIR, XX 	\\ %
HD 294271    		&  3.9$^{+0.8}_{-0.6}$	&		\\ %
HD 37525 AB  		&  3.6$^{+0.7}_{-0.6}$	& binary, XX 	\\ %
HD 294272 A  		&  3.2$\pm$0.5	  	&		\\ %
HD 294272 B  		&  3.0$\pm$0.5  	&		\\ %
HD 37333     		&  3.0$^{+0.6}_{-0.4}$	&		\\ %
HD 37564     		&  3.1$^{+0.6}_{-0.4}$	& 		\\ %
$\sigma$ Ori C 		&  2.7$\pm$0.4  	&		\\ %
V1147 Ori    		&  3.5$^{+0.4}_{-0.3}$ 	& XX 		\\ %
HD 37686     		&  2.3$^{+0.4}_{-0.1}$	&		\\ %
HD 37545     		&  2.2$^{+0.3}_{-0.1}$	&		\\ %
HD 294275    		&  2.1$^{+0.3}_{-0.1}$	&		\\ %
HD 294297 		&  1.7$^{+0.2}_{-0.1}$	& $\mu$ 	\\ %
HD 294268		&  1.6$^{+0.2}_{-0.1}$	& mIR 		\\ %
RX J0539.6--0242 AB	&  1.8$\pm$0.1  	& SB, XX	\\ %
HD 294279    		&  1.6$\pm$0.1  	&		\\ %
HD 294273 		&  1.5$\pm$0.1  	&		\\ %
StHA 50  		&  1.4$\pm$0.1  	&  		\\ %
HD 294301	    	&  1.5$\pm$0.1  	& prob. 	\\ %
GSC 04771--00962	&  1.4$\pm$0.1  	& prob. 	\\ %
HD 294298    		&  1.5$\pm$0.1  	& XX		\\ %
4771--0950		&  1.3$\pm$0.1  	&		\\ %
2E 0535.4--0241		&  1.5$\pm$0.1  	& XX		\\ %
TYC 4770 1261 1		&  1.3$\pm$0.1  	& prob. 	\\ %
TYC 4770 1129 1		&  1.3$\pm$0.1  	& prob. 	\\ %
SO120532	      	&  1.2$\pm$0.1  	&		\\ %
            \noalign{\smallskip}
            \hline
         \end{tabular}
     $$ 
\begin{list}{}{}
\item[$^{a}$] Remarks:
``mIR'': IRAS source;
``XX'': moderate or strong X-ray emitter;
``prob.'': probable association member star without detected youth features.
The rest are described in the text.
\end{list}
   \end{table}

There are three bright stars ($V <$ 9.5\,mag) that are 0.1--0.2\,mag redder than
the isochrones and the sequence defined by the other cluster members in the
colour-magnitude diagram.
They are $\sigma$ Ori B (which atmosphere could being suffering the effects of
the O9.5V primary at only $\sim$ 90\,AU), $\sigma$ Ori E (a helium rich star --
i.e. the used models may be not applicable) and HD 37564 (which may have a disc
-- see Section \ref{discos}). 
The masses derived in this work for these objects must be cautiously treated.
The rest of the bright stars also lie slightly to the red of the isochrones in
the colour-magnitude diagram in Figure \ref{F04}.  
Since in the high-mass domain of the colour-magnitude diagram in Fig.
\ref{F04} there is almost no sensitivity to the age of the cluster, this
redshift favours the nearest possible heliocentric distances to $\sigma$
Orionis. 
It implies thus that the cluster could be older than usually assumed. 
This fact of the greatest importance for the derivation of the mass function in
the cluster substellar domain down to planetary masses (B\'ejar et al.
2001; Gonz\'alez-Garc\'{\i}a et al. 2006; Caballero et al. 2007a), since
the theoretical masses from the models strongly depend on the age. 
An age older than expected for $\sigma$ Orionis, at about 10\,Ma,
would lead dozens currently considered brown dwarfs to be very
low-mass stars, and several planetary-mass objects to be really 
brown dwarfs.

\subsection{Disc harbours and X-ray emitters}
\label{discos}

Some of the young stars in Table \ref{estrellas_masas} display characteristics
that are present in T Tauri and Herbig Ae/Be stars.
In the standard scenario, these stars possess discs, which are
responsible of the flux excess in the mid and/or near infrared.
The in-falling material from the disc onto the stellar surface produces the X-ray
emission and the strong broad H$\alpha$ emission. 
The columns of the accreting mass hitting the atmosphere can produce
photometric variability and flux excess in the blue optical as well
(Bertout et al. 1989; Appenzeller \& Mundt 1989; K\"onigl 1991; Hillenbrand et
al. 1992).

\begin{figure}
\centering
\includegraphics[width=0.49\textwidth]{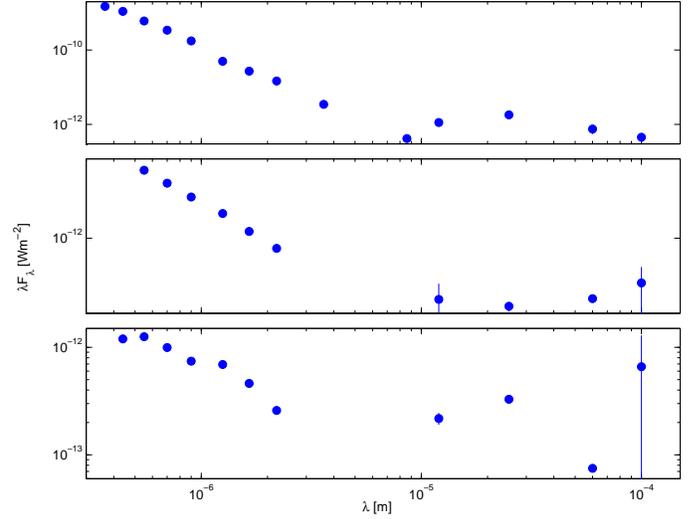}
\caption{Spectral energy distributions of $\sigma$ Ori A+B+IRS1 (top), HD
37699 (middle) and HD 294268 (bottom).
Vertical scales are different.
}
\label{F07}
\end{figure}
%

   \begin{table}
      \caption[]{Young stars with probable discs$^a$.} 
         \label{estrellas_discos}
     $$ 
         \begin{tabular}{l c r}
            \hline
            \hline
            \noalign{\smallskip}
Name 			& Sp. type	& Remarks \\  
            \noalign{\smallskip}
            \hline
            \noalign{\smallskip}
HD 37699 		& B5Vn		& IRAS source \\ %
HD 37564 		& A8V:		& red $B-V$ \\ %
HD 294268 		& F5		& IRAS source \\ %
RX J0539.6--0242 AB 	& G5--K0	& $V_T-K_{\rm s} >$ 2.0\,mag \\ %
StHA 50 		& Be		& blue $B-V$ \\ %
HD 294298 		& G0:		& $V_T-K_{\rm s} >$ 2.0\,mag \\ %
2E 0535.4--0241 	& 		& $V_T-K_{\rm s} >$ 2.0\,mag \\ %
            \noalign{\smallskip}
            \hline
         \end{tabular}
     $$ 
\begin{list}{}{}
\item[$^{a}$] Discarding $\sigma$ Ori IRS1.
\item[$^{b}$] Weak-line T Tauri binary (Alcal\'a et al. 1996).
\end{list}
   \end{table}

Among the investigated stars, three were detected by the {\em IRAS} space
telescope (Table \ref{estrellas_mIR} in Section \ref{iras}).
I have built their spectral energy distributions (SEDs) from the blue optical to
100\,$\mu$m ($(U)BVRIJHK_{\rm s}(L'[8.6])[12][25][60][100]$), using the data
presented here and from Lee (1968) and van Loon \& Oliveira (2004).
There is high-resolution (sub-arcsec) imaging for only the $\sigma$ Ori A+B+IRS1
system.
The mid-infrared source $\sigma$ Ori IRS1, at 3.3\,arcsec northeast to the OB
system $\sigma$ Ori AB, has been resolved  at 6--20\,$\mu$m (van Loon \& 
Oliveira 2003), in X rays (Sanz-Forcada et al. 2004), in the $J$ and $H$
bands (Caballero 2005) and in the optical (Rebolo et al. in prep.). 
It is the responsible of the IRAS emission coming from the triple system.
I propose that $\sigma$ Ori IRS1 is a low-mass K:-type star surrounded by a
thick envelope that loses its outer atmospheric layers due to the strong
ultraviolet radiation field and stellar wind coming from the O9.5V+B0.5V system
at only $\sim$1000\,AU (see also van Loon \&  Oliveira 2003).
The other two stars with IRAS emission, HD 37699 and HD 294268, have masses of
4.4$^{+0.8}_{-0.7}$ and 1.6$^{+0.2}_{-0.1}$\,M$_\odot$, respectively, which make
them to be suitable candidate Herbig Ae/Be stars.
Besides, according to Bertout (1989), the more massive T Tauri stars rotate
significantly faster than less massive ones.
The large $v \sin{i}$ derived from the ``nebulous'' spectral type of HD 37699
(Neubauer 1943) points out that this relation may be extrapolated to Ae/Be
stars. 
The disc surrounding HD 37699 seems also to be cooler than those of $\sigma$ Ori
IRS1 and HD 294268, since the flux density at 60\,$\mu$m is larger than at
25\,$\mu$m, to the contrary to what happens in the other two.
The large uncertainties in the {\em IRAS} fluxes prevent to derive additional
properties of the discs.
The abrupt step in the SED of HD 294268 between the $BVRI$-band data (Lee 1968)
and the 2MASS $JHK_{\rm s}$-band data (Cutri et al. 2003) suggests a secular
large-amplitude photometric variability.

\begin{figure}
\centering
\includegraphics[width=0.53\textwidth]{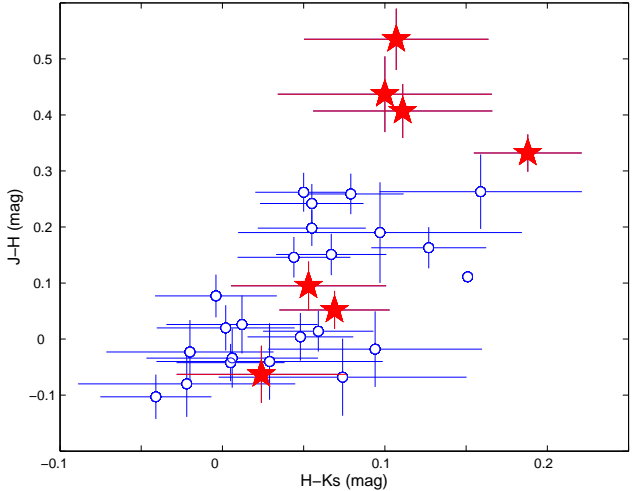} 
\caption{$J-H$ vs. $H-K_{\rm s}$ colour-colour diagram of the brightest
stars of the $\sigma$ Orionis cluster.
Stars with and without disc are marked with filled stars and open circles,
respectively.
The error bars of $\sigma$ Ori AB have not been drawn for clarity.
}
\label{F06.5}
\end{figure}

There are five additional candidate star systems with probable discs.
They are shown, together with HD 37699 and HD 294268, in Table
\ref{estrellas_discos}.
Three of them are the strong X-ray emitters with colours $V_T-K_{\rm s} >$
2.0\,mag (Section \ref{KandM}).
The remaining probable disc harbours are HD 37564 and StHA 50.
The former is an A8:-type star with moderate X-ray emission and relatively red
$B-V$ and $V_T-K_{\rm s}$ colours.
StHa 50, for which there is only a few data available (it is a Be-type
emission-line object with $V \approx$ 11.4\,mag -- Stephenson 1986; Downes \&
Keyes 1988), deviates at the level of 3 $\sigma$ to the blue in the $V$ vs.
$B-V$ colour-magnitude diagram in Figure \ref{F04} respect to the sequence
defined by the other cluster members.
IRAC {\em Spitzer} Space Telescope data between 4.5 and 8.0\,$\mu$m supports the
hypothesis of discs surrounding both HD 37564 and StHA 50 (Caballero et al.
in prep.).
{As a matter of fact, both of them and HD 37699, the bluest star with a disc
in $\sigma$ Orionis, have no appreciable near-infrared flux excess in the $J-H$
vs. $H-K_{\rm s}$ colour-colour diagram shown in Fig. \ref{F06.5}.
Their discs are probably cooler than those surrounding the remaining stars in
Table \ref{estrellas_discos}. 

   \begin{table}
      \caption[]{Frequency of discs in different mass intervals in the Ori
      OB 1b Association and in the $\sigma$ Orionis cluster.} 
         \label{discfrequency}
     $$ 
         \begin{tabular}{lcr}
            \hline
            \hline
            \noalign{\smallskip}
Frequency 	& Reference			& Method$^{a}$	\\  
(\%) 		& 				& 		\\   
            \noalign{\smallskip}
            \hline
            \noalign{\smallskip}
17$^{+18}_{-17}$ & Jawawardhana et al. (2003)	& Ja03	\\  
5--12 		& Barrado y Navascu\'es et al. (2003)	& ByN03	\\  
46--54 		& Oliveira et al. (2004)	& Ol04	\\  
$\sim$25 	& Caballero et al. (2004)	& Ca04	\\  
5--7 		& Scholz \& Eisl\"offel (2004)	& SE04	\\  
17.4$\pm$3.6 	& Hern\'andez et al. (2005)	& He05	\\  
10$\pm$5	& Kenyon et al. (2005)		& Ke05	\\
46$^{+16}_{-13}$ & Caballero (2005)		& Ca05	\\  
33$\pm$6 	& Oliveira et al. (2006)	& Ol06	\\  
33$\pm$4 	& Caballero et al. (2006b)	& Ca06b	\\  
47$\pm$14 	& Caballero et al. (2007a)	& Ca07a	\\  
30$\pm$20 	& Caballero et al. (2007b)	& Ca07b	\\  
32$\pm$14 	& This work			& 	\\  
            \noalign{\smallskip}
            \hline
         \end{tabular}
     $$ 
\begin{list}{}{}
\item[$^{a}$]
(Ja03): very low-mass stars in $\sigma$ Orionis with $L'$ excess; 
(ByN03): intermediate- and low-mass stars and high-mass brown dwarfs in $\sigma$
Orionis with $K_{\rm s}$ excess (minimum frequency); 
(Ol04): low-mass stars and high-mass brown dwarfs in $\sigma$ Orionis with
$K_{\rm s}L'$ excess; 
(Ca04): brown dwarfs in $\sigma$ Orionis with $K_{\rm s}$ excess, H$\alpha$
emission and/or large-amplitude photometric variability in the optical; 
(SE04): very low-mass stars and high-mass brown dwarfs in $\sigma$ Orionis with
high-level variability and/or strong H$\alpha$ and Ca-triplet emission; 
(He05): bright stars in the Ori OB 1bc (super-)association using an $HK$
criterion;  
(Ke05): low-mass stars and high-mass brown dwarfs in $\sigma$ Orionis with
broadened H$\alpha$ emission;  
(Ca05): K and M stars in $\sigma$ Orionis with broadened H$\alpha$ emission; 
(Ol06): low-mass stars and high-mass brown dwarfs in $\sigma$ Orionis with
$K_{\rm s}L'$ excess; 
(Ca06b): 1.0--0.035\,M$_\odot$ stars and high-mass brown dwarfs in $\sigma$
Orionis with excess at 8.0\,$\mu$m;
(Ca07a): brown dwarfs in $\sigma$ Orionis with excess at 5.8--8.0\,$\mu$m;
(Ca07b): isolated planetary-mass objects in $\sigma$ Orionis with excess at
5.8\,$\mu$m.
\end{list}
   \end{table}

The frequency of discs surrounding young stars in the magnitud interval
7.5\,mag $\lesssim V \lesssim$ 11.0\,mag and in the direction of $\sigma$ Ori is
32$\pm$14\,\%. 
This value is of the order of other disc frequencies measured in the area
in different mass intervals (compiled in Table \ref{discfrequency}). 
The frequencies derived from the detection of flux excess red-wards 
5\,$\mu$m are in general larger than the disc frequencies based in only 
$JHK_{\rm s}$ photometry. 
Since the various disc frequencies quoted have been assessed heterogeneously
with different techniques, I do not claim any evidence for a mass-dependence of
the disc frequency in $\sigma$ Orionis. 

The strongest X-ray emitters are $\sigma$ Ori A+B+IRS1 and $\sigma$ Ori E.
The $\sigma$ Ori A+B+IRS1 system was resolved in the AB and IRS1 components
with {HRI-C}/{\em Chandra} Space Telescope observations.
The bulk of the X-ray flux comes from the OB pair (Sanz-Forcada et al. 2004).
Other moderate emitters are HD 37699, HD 37525 AB, V1147 Ori
and the three stars with $V_T-K_{\rm s} >$ 2.0\,mag in Table
\ref{estrellas_discos}.
Four bright stars out of the eight moderate and strong X-ray emitters seem to
harbour surrounding discs, which supports the scenario of T Tauri and Ae/Be
stars. 

The Figure \ref{F08} shows the variation between the years 1991 and 1994 of the
background subtracted count rate in the energy interval 0.24--2.0\,keV of
the flux of the young stars found in WGACAT.
Several targets display flares and variability of the X-ray emission in time
scales from days to years.
For example, the flux from $\sigma$ Ori AB (top light-curve in Figure
\ref{F08}) raised from 0.257$\pm$0.005 to 0.452$\pm$0.011\,ct\,s$^{-1}$ between
1993 Feb. 26 and 28, while the flux from V1147 Ori decreased in a factor 5 with
respect to an hypothetical quiescent level at $\sim$0.009\,ct\,s$^{-1}$ in 
observations spaced out about one year.
Both $\sigma$ Ori AB and V1147 Ori do not apparently possess discs, so the
X-ray variability is probably due to stellar winds and chromo- or
magnetospheric effects, respectively.

\begin{figure}
\centering
\includegraphics[width=0.49\textwidth]{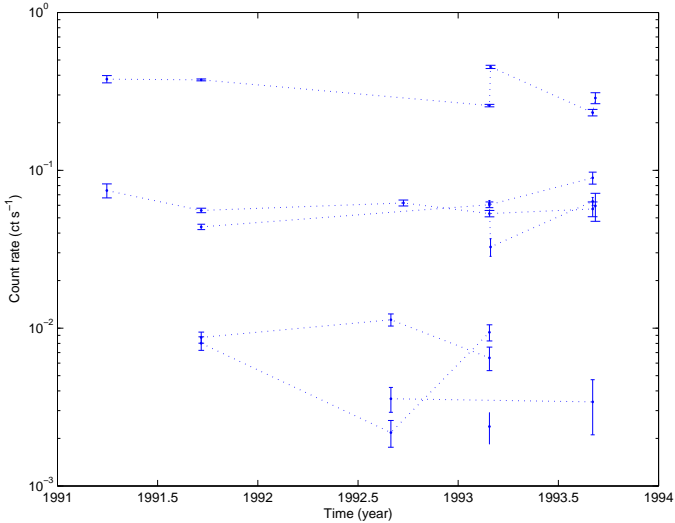}
\caption{X-ray light curves of the brightest $\sigma$ Orionis stars in the
{\em ROSAT} WGACAT catalogue.}
\label{F08}
\end{figure}

\subsection{Different stellar populations surrounding $\sigma$ Ori}
\label{different}

There are five Hipparcos stars in the investigated area, which HIP
identifications and parallax determinations are provided in Table
\ref{estrellas_hipparcos}.
The data may suggest that the stars are located at different heliocentric
distances. 
However, the Hipparcos parallax measurements at more than 100\,pc are quite not
reliable. 
Furthermore, in at least three cases the measurements are affected by
multiplicity.
It is expected that the {\em GAIA} mission will be able to disentangle distinct
populations in the Ori OB1 Association located at different heliocentric
distances.

Meanwhile, it is possible to disentangle thanks to the knowledge of the spatial
velocities of the young stars.
Previous determinations of the radial velocity of $\sigma$ Ori AB and the
mean radial velocity of the $\sigma$ Orionis cluster ranged in a 
relatively narrow interval between +25 and +35\,km\,s$^{-1}$ (Frost et al. 
1926; Heard 1949; Wilson 1953; Evans 1967; Bolton 1974; Cruz-Gonz\'alez et al.
1974; Conti et al. 1977; Morrell \& Levato 1991; Zapatero Osorio et al. 2002;
Muzerolle et al. 2003; Burningham et al. 2005; Kenyon et al. 2005; Caballero
2005). 
The two recent studies by Jeffries et al. (2006) and Sacco et al. (2006)
are in agreement that the heliocentric radial velocity (V$_r$)}is about
+31.0$\pm$0.5\,km\,s$^{-1}$ with a dispersion less than 8\,km\,s$^{-1}$ (group 2
of Jeffries et al. 2006).
I have compiled all the published values of V$_r$ of the investigated stars.
They are provided in the last column of Table \ref{estrellas_dos}.
Among the nine stars with measured V$_r$, there is a non-member star (HD
294269, already discussed in Sections \ref{optical} and \ref{propermotion}).
The remaining stars have been classified in this work as members and possible
members of the Ori OB 1b Association. 
$\sigma$ Ori AB, D and E obviously display V$_r$ similar to the mean value of
Jeffries et al. (2006).
The same occurs for the classical T Tauri star RX J0539.6--0242 AB, at
$\sim$15\,arcmin to the centre of the cluster. 
However, the measurements at two different epochs of the radial velocity of the
Ae/Be star HD 37699 differ from the mean radial velocity
(+14.8$\pm$2.2\,km\,s$^{-1}$, Neubauer 1943; +15\,km\,s$^{-1}$, Duflot et al.
1995). 
The radial velocity of HD 294298 is also quite discordant, V$_r$ =
+11.5\,km\,s$^{-1}$.
Both HD 37699 and HD 294298 are closely located (in the northeast outer part
of the studied area, in the vicinity of the Horsehead and Alnitak), have similar
proper motions (identical within the error bars) and possess discs.
HD 294297 (V$_r$ = +25\,km\,s$^{-1}$) and HD 294268 (V$_r$ =
+20\,km\,s$^{-1}$) apparently display different V$_r$ from the mean value of
$\sigma$ Orionis as well.   

\begin{figure}
\centering
\includegraphics[width=0.49\textwidth]{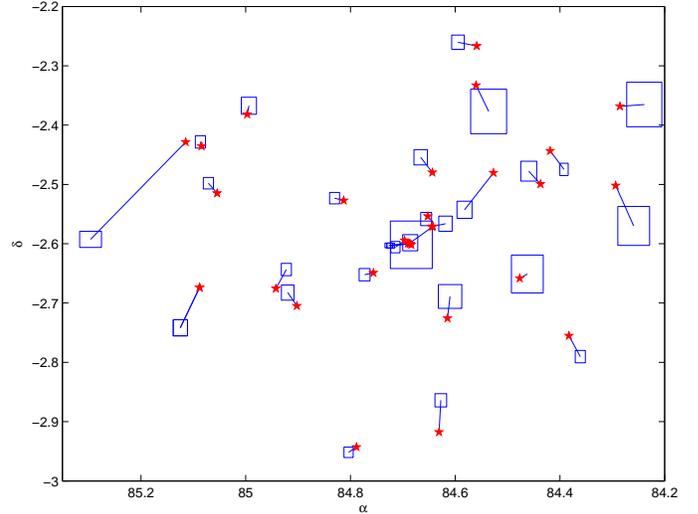}
\caption{Current (filled stars) and future (boxes) positions of the young stars
and candidates. 
Sizes of the boxes are proportional to the uncertainties.}  
\label{F09}
\end{figure}

The eastern star in this sample, also close to the Horsehead Nebula, is HD
294297 (with discordant radial velocity and lithium in absorption; Section
\ref{tycho-1}). 
Its proper motion is $\gtrsim$ 30\,mas\,a$^{-1}$ larger than that of $\sigma$
Ori AB and similar to those of probable foreground stars with large tangential
velocities (see also Zapatero Osorio et al. 2006).
The expected positions in the future of HD 294297 and of the other 29 young
stars are shown in Figure \ref{F09}. 
While most of the stars keep their positions in the sky or float
around, in the same period of time HD 294297 would have crossed the Horsehead
Nebula and be located out of the 30\,arcmin-radius area.

%

   \begin{table}
      \caption[]{Young stars in the Hipparcos catalogue.} 
         \label{estrellas_hipparcos}
     $$ 
         \begin{tabular}{lccr}
            \hline
            \hline
            \noalign{\smallskip}
Name 			& HIP	& $\pi \pm \delta \pi$	& Sp.\\  
 			& 	& [mas]			& type \\  
            \noalign{\smallskip}
            \hline
            \noalign{\smallskip}
$\sigma$ Ori A+B+IRS1 	& 26549	& 2.84$\pm$0.91 	& O9.5V+... \\ %
$\sigma$ Ori D  	& 26551 & 2.84$\pm$0.91 	& B2V \\ %
HD 37699  	  	& 26694	& 0.63$\pm$1.01 	& B5Vn \\ %
HD 37525 AB 	  	& 26579	& 2.55$\pm$0.99 	& B5Vp \\ %
HD 37333  	  	& 26438	& 2.52$\pm$1.08 	& A1Va \\ %
            \noalign{\smallskip}
            \hline
         \end{tabular}
     $$ 
   \end{table}

From the existence of such young stars with peculiar spatial velocities, it can
be derived that:   
($i$) the astrometric criterion used in Section \ref{propermotion} to discard
stars as possible cluster members could be too conservative (but necessary to
minimize the number of contaminant foreground stars in the list of candidate
young stars); and
($ii$) there are young stars in the line of sight of the $\sigma$ Orionis
cluster probably originated in a different star-forming region in the Ori OB 1b
Association.
The second item supports the hypothesis of the existence of at least two
spatially superimposed components around $\sigma$ Ori exposed by Jeffries et al.
(2006).
They provided an excellent review of some possible consequences of a population
with mixed age and distance.
However, given that their spectroscopic studied covered a large area, they
investigated young stars that are out of the 30\,arcmin-radius and closer to the
extremely bright O9.5Ib-type Alnitak star than to $\sigma$ Ori AB itself.
Three out of the proper-motion outlier young stars in this Section are
located also in the direction of Alnitak, very close to the Horsehead
Nebula. 
What we see may be, in reality, an overlapping of sub-associations linked to
different, but relatively close, star-forming regions.
This was firstly noted by B\'ejar et al. (2004a), who proposed that an
overdensity of photometric cluster members to the north-east of $\sigma$ Orionis
could be associated to an hypothetical cluster surrounding Alnitak.
They also studied the radial distribution of low-mass stars and brown
dwarfs in a region around $\sigma$ Ori, and found that the characteristic
radius of the surface density, $\rho(r)$, was $r_0$ = 8.8$\pm$0.6\,arcmin (where
$\rho(r) = \rho_0 e^{-r/r_0}$ and $r$ is the separation to the cluster centre).
Since the spatial distribution of stars (split in different mass intervals) and
substellar objects in the cluster are identical (Caballero 2006),
the spatial density of stellar members at $\sim$26\,arcmin ($r = 3 r_0$) is
about 5\,\% of the central density.
Similar results were found by Sherry et al. (2004).
Therefore, the contamination by young stars that do not belong to the $\sigma$
Orionis cluster in wide searches like this may be significative, especially at 
large separations to the centre.
In particular, the four stars with abnormal spatial velocities are located at
21--28\,arcmin to $\sigma$ Ori AB, which supports this hypothesis.
The only way to make much headway against these problems is to obtain precise
V$_r$ over a wider area, quantify the number of young contaminants and determine
if they belong to any of the sub-groups presented by Jeffries et al. (2006) or
to new ones. 
Multi-epoch observations are mandatory, since bona-fide $\sigma$ Orionis member
spectroscopic binaries could appear as V$_r$ outliers and vice versa.
Besides, the binary frequency in the higher mass domain of the cluster seems to
be as large as $\sim$50\,\% (Caballero 2005).

\subsection{Total mass and mass function from 1.1 to 24\,M$_\odot$} 
\label{massspectrum}

The masses derived for the stars of this sample are about 20\,\% less than
previous determinations in the literature (e.g. Hunger et al. 1989 and
Hern\'andez et al. 2005). 
The list of the twelve stars of $\sigma$ Orionis more massive than
3\,M$_\odot$ basically matches that prepared by Sherry et al. (2004).
They derived $\sim$ 100\,M$_\odot$ as the mass exclusively contained in them.
However, in the present work, the stars with M $>$ 3\,M$_\odot$ account for
only $\sim$ 64\,M$_\odot$, which may lead to a cluster total mass
significantly lower than previously estimated (225$\pm$30\,M$_\odot$; Sherry et
al. 2004).
Simply by scaling the correction factor down to lower masses, the cluster total
mass could be at the level of only $\sim$150\,M$_\odot$.  
The total mass could be even lower accounting for probable young
contaminants.

The total mass of the stars in this sample (probably all the $\sigma$ Orionis
stars more massive than $\sim$1.2\,M$_\odot$ plus some young contaminant)
is 100$^{+20}_{-19}$\,M$_\odot$. 
Out of it, 46\,\% is contained in the $\sigma$ Ori A+B+C+D+E system, and up to
29\,\% only in the two most massive stars in the cluster, $\sigma$ Ori A and B.
This fact indicates that the cluster presents a well-defined cluster centre, in
contrast to other extended star-forming regions, as 
Upper Scorpius (de Zeeuw et al. 1999; Preibisch \& Zinnecker 1999).
Discarding the three stars discussed in the Section \ref{different}, the
cluster barycentre is located at $\sim$ 1\,arcmin to the west of $\sigma$ Ori
AB. 
The mean tangential velocity of the cluster centre of mass is 
($\mu_\alpha \cos{\delta}$, $\mu_\delta$) = 
(+2.2$\pm$1.2, {--0.5$\pm$1.0})\,mas\,a$^{-1}$,
which is also quite similar to the tangential velocity of $\sigma$ Ori AB 
(the result rejecting the three stars is identical within the error
bars)\footnote{The standard deviation with respect to the mean proper motion (no
considering HD 37699, HD 294298 and HD 294297) is 5.1\,mas\,a$^{-1}$, which
transforms into a tangential velocity dispersion of 8.4\,km\,s$^{-1}$ at the
heliocentric distance of 360\,pc.}.

\begin{figure}
\centering
\includegraphics[width=0.49\textwidth]{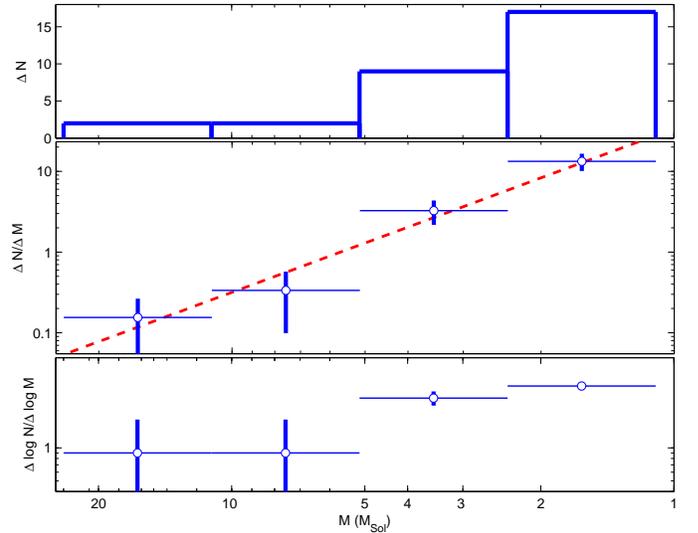}
\caption{The cumulative number of stars per mass interval (top), the mass
spectrum (middle) and the mass function (bottom) in $\sigma$ Orionis in the mass
interval 1.1--24\,M$_\odot$.
A distance of 360\,pc and an age of 3\,Ma have been used. 
In the middle window, the dashed (red) line marks the fit of the mass spectrum
($\alpha$ = +2.0).
}
\label{F06}
\end{figure}

Last, but not less important, I have also computed the mass spectrum in $\sigma$
Orionis in the mass interval 1.1--24\,M$_\odot$ from the masses shown in Table
\ref{estrellas_masas}. 
The Figure \ref{F06} displays the cummulative number of stars per mass interval
($\Delta {\rm N}$), the mass spectrum 
($\Delta {\rm N} / \Delta {\rm N} = a {\rm M}^\gamma$; {$\alpha \equiv -\gamma$})
and the mass function 
($\Delta \log{\rm N} / \Delta \log{\rm N} = A {\rm M}^\Gamma$; {$\alpha \equiv 1
-\Gamma$}) in the cluster. 
The value of the $\alpha$ index from the fit of the mass spectrum for 3\,Ma and
360\,pc gave  $\alpha$ = +2.0 (the fit rejecting the four stars in the
Section \ref{different} offers the same result).
Considering the very conservative uncertainties in the age and heliocentric
distance, and repeating the fit for the less probable cases (1\,Ma and 300\,pc,
10\,Ma and 430\,pc), then the value of the index in the mass spectrum is
$\alpha$ = +2.0$^{+0.2}_{-0.1}$.
This value is slightly smaller than, but quite consistent with, the $\alpha$
value for massive and solar-like stars ($\alpha \approx$ +2.3; Salpeter 1955).
It may be due to the incompleteness of the search below 1.3\,M$_\odot$
($V_T \gtrsim$ 11.5\,mag).
However, the measured $\alpha$ nicely extrapolates the Sherry et al. (2004)
studies in $\sigma$ Orionis. 
They found that the cluster mass function in the mass interval
0.2--1.0\,M$_\odot$ is also consistent with that of the field stars in the same
mass interval by Kroupa (2001, 2002).
Therefore, the index of the mass spectrum in $\sigma$ Orionis keeps roughly
constant at $\alpha \sim$ 2 from 24 to 0.2\,M$_\odot$, where there is a turning 
point (first mentioned by Caballero 2006).
The $\alpha$ index below 0.2\,M$_\odot$ is established to be in the range
+0.6--0.8 (B\'ejar et al. 2001; Gonz\'alez-Garc\'{\i}a et al. 2006).
Tej et al. (2002), who measured the mass spectrum in an interval covering both
sides of the turning point (0.5--0.08\,M$_\odot$), obtained an intermediate
value of $\alpha$ = +1.2$\pm$0.1. 
Then, the $\sigma$ Orionis cluster has become an excellent laboratory to study
the mass function from the most massive stars down to planetary-mass objects.

\section{Summary}

This work is devoted to identify and characterize the brightest stars of the
$\sigma$ Orionis cluster.
The results presented here will complement searches of low-mass stars,
brown dwarfs and planetary-mass objects in the cluster.

From a correlation between the Tycho-2 and 2MASS catalogues, and using
additional information from X-ray and mid-infrared space missions and from the
literature, I have classified the brightest stars in a circular area of radius
30\,arcmin centred in $\sigma$ Ori AB in:  
($i$) members of the Ori OB 1b Association with youth features (26);
($ii$) candidate young stars (4); and
($iii$) probable foreground stars with red $V_T-K_{\rm s}$ colours and/or large
tangential velocities (16).

The comparison of the sequence of the young stars with the lowest photometric
errors and evolutionary tracks in the $V$ vs. $B-V$ diagram suggests that
$\sigma$ Orionis could be closer and older than previously considered.

I have derived the masses from theoretical $M_V$-mass relations for the young
stars and candidate young stars and computed for the first time the mass
spectrum of the cluster between 1.1 and 24\,M$\odot$.
The $\alpha$ index from the fit of the mass spectrum is +2.0$^{+0.2}_{-0.1}$,
which is quite similar to the Salpeter's index.
The cluster total mass could be as low as $\sim$ 150\,M$\odot$.

From the flux excess in the mid and/or near infrared, the abnormal position in
the $V$ vs. $B-V$ diagram and the strong or moderate X-ray emission, I have
derived the existence of discs surrounding seven stars, four out of
which are new. 
The frequency of discs in the magnitude interval 7.5\,mag $\lesssim V \lesssim$
11.0\,mag is 32$\pm$14\,\%. 

Finally, the existence of some stars with abnormal tangential and radial
velocities supports the hypothesis of overlapping of stellar populations of
different ages and kinematics in the direction of the $\sigma$ Orionis cluster.

\appendix 

\section{Stellar data}

Table \ref{estrellas_uno} provides the photometric data for the 41 stars in
the Tycho-2/2MASS correlation (above the horizontal line) and the five missing
stars in the Tycho-2/2MASS correlation (below the horizontal line), ordered by
$V_T$-band magnitude.
It is given the name of each star (if not available, then the Tycho
identification is provided), the $B_T$- and $V_T$-band magnitudes from the
Tycho-2 catalogue and the $J$-, $H$- and $K_{\rm s}$-band magnitudes from the
2MASS catalogue. 

Table \ref{estrellas_dos} provides the astrometric data and the spectral
types in the same manner as in Table \ref{estrellas_uno}.
It is given the name of each star, right ascension and declination from the
Tycho-2 catalogue (FK5, Equinox=J2000.0, Epoch=J2000.000, proper motions taken
into account), proper motions in right ascension and in declination from the
Tycho-2 catalogue and spectral types from the literature (references in the
text).

\begin{acknowledgements}

I thank C. Bertout, D. J. Lennon, T. J. Mahoney, V. J. S\'anchez B\'ejar, E.
Solano, M. R. Zapatero Osorio and, especially, the referee, R. D.
Jeffries, for providing me helpful comments at different stages of the 
writing.  
Partial financial support was provided by the Spanish Ministerio de
Ciencia y Tecnolog\'{\i}a project AYA2004--00253 of the Plan Nacional de
Astronom\'{\i}a y Astrof\'{\i}sica. 
This research has made use of the SIMBAD database, operated at CDS,
Strasbourg, France.  
This publication makes use of data products from the Two Micron All Sky
Survey, which is a joint project of the University of Massachusetts and
the Infrared Processing and Analysis Center/California Institute of
Technology, funded by the National Aeronautics and Space
Administration and the National Science Foundation.

\end{acknowledgements}

\include{6652Ts}

\end{document}

%% file: 6652Ts.tex
   \begin{table*}
      \caption[]{Optical and near-infrared photometry of all the stars.}
         \label{estrellas_uno}
     $$ 
         \begin{tabular}{lccccccccr}
            \hline
            \hline
            \noalign{\smallskip}
Name				& $B_T \pm \delta B_T$ & $V_T \pm \delta V_T$ & $J \pm \delta J$ & $H \pm \delta H$ &$K_{\rm s} \pm \delta K_{\rm s}$ \\  
   				& [mag] & [mag] & [mag] & [mag] & [mag] \\  
            \noalign{\smallskip}
            \hline
            \noalign{\smallskip}
$\sigma$ Ori A+B+IRS1 		&  3.521 0.014 &  3.763  0.009  &  4.752  0.260 &  4.641  0.252 &  4.490  0.016 \\
\object{$\sigma$ Ori E} 	&  6.236 0.014 &  6.344  0.010  &  6.974  0.026 &  6.954  0.031 &  6.952  0.029 \\
\object{HD 37699}     		&  7.457 0.015 &  7.590  0.011  &  7.841  0.029 &  7.904  0.042 &  7.880  0.031 \\
\object{HD 294271}    		&  7.757 0.016 &  7.856  0.011  &  8.100  0.021 &  8.180  0.055 &  8.202  0.038 \\
\object{HD 37525} AB  		&  7.952 0.016 &  8.058  0.012  &  8.131  0.030 &  8.105  0.042 &  8.093  0.020 \\
\object{HD 294272 A}  		&  8.402 0.017 &  8.389  0.015  &  8.346  0.026 &  8.380  0.046 &  8.374  0.026 \\
\object{HD 294272 B}$^{a}$  	&  8.544 0.018 &  8.554  0.015  &  8.779  0.021 &  8.819  0.065 &  8.790  0.025 \\
\object{HD 37333}     		&  8.616 0.017 &  8.508  0.013  &  8.413  0.023 &  8.481  0.065 &  8.407  0.040 \\
\object{HD 37564}     		&  8.699 0.017 &  8.443  0.014  &  7.976  0.021 &  7.881  0.038 &  7.828  0.029 \\
\object{V1147 Ori}    		&  9.065 0.018 &  8.992  0.016  &  8.876  0.024 &  8.894  0.063 &  8.800  0.019 \\
\object{HD 37686}     		&  9.210 0.019 &  9.186  0.018  &  9.207  0.035 &  9.203  0.024 &  9.155  0.022 \\
\object{HD 37545}     		&  9.263 0.018 &  9.277  0.017  &  9.261  0.023 &  9.303  0.024 &  9.298  0.023 \\
\object{HD 294275}    		&  9.504 0.020 &  9.431  0.019  &  9.256  0.028 &  9.242  0.023 &  9.183  0.025 \\
\object{HD 294268}		& 10.966 0.049 & 10.507  0.043  &  9.393  0.023 &  9.061  0.024 &  8.873  0.023 \\
\object{HD 294307}    		& 11.091 0.054 & 10.552  0.046  &  9.439  0.026 &  9.177  0.023 &  9.127  0.019 \\
\object{RX J0539.6--0242} AB$^{b}$&11.103 0.055& 10.195  0.036  &  8.462  0.027 &  8.055  0.040 &  7.944  0.038 \\
\object{HD 294279}    		& 11.131 0.054 & 10.685  0.053  &  9.873  0.028 &  9.727  0.023 &  9.683  0.026 \\
\object{EM* StHA 50}  		& 11.219 0.059 & 11.275  0.089  & 10.666  0.024 & 10.614  0.024 & 10.545  0.024 \\
\object{GSC 04771--00621}	& 11.231 0.059 & 10.920  0.063  &  9.796  0.027 &  9.600  0.023 &  9.503  0.027 \\
\object{HD 294277}    		& 11.416 0.063 &  9.737  0.024  &  6.773  0.017 &  6.066  0.034 &  5.861  0.021 \\
\object{HD 294278}    		& 11.449 0.067 &  9.941  0.029  &  7.592  0.023 &  7.034  0.053 &  6.855  0.033 \\
\object{HD 294270}    		& 11.507 0.067 & 10.960  0.062  &  9.780  0.024 &  9.511  0.025 &  9.472  0.025 \\
\object{HD 294301}    		& 11.541 0.079 & 11.139  0.084  & 10.210  0.028 & 10.059  0.024 &  9.992  0.024 \\
\object{HD 294269}    		& 11.597 0.076 & 10.795  0.011  &  9.178  0.035 &  8.721  0.033 &  8.634  0.027 \\
\object{GSC 04771--00962}	& 11.608 0.073 & 11.101  0.075  & 10.202  0.026 &  9.960  0.023 &  9.905  0.022 \\
\object{HD 294298}    		& 11.646 0.086 & 10.975  0.073  &  9.339  0.029 &  8.902  0.061 &  8.802  0.025 \\
\object{HD 294274}    		& 11.662 0.071 & 10.758  0.056  &  9.428  0.027 &  9.069  0.023 &  8.985  0.025 \\
\object{[W96] 4771--0950}$^{c}$	& 11.807 0.086 & 11.429  0.099  & 10.088  0.027 &  9.829  0.024 &  9.750  0.022 \\
\object{HD 294280}    		& 11.893 0.091 &  9.865  0.027  &  6.429  0.019 &  5.580  0.031 &  5.340  0.024 \\ 
\object{TYC 4771 720 1}		& 11.944 0.105 & 11.787  0.152  & 10.269  0.028 &  9.885  0.023 &  9.813  0.021 \\
\object{2E 0535.4--0241}	& 12.021 0.117 & 11.015  0.069  &  9.255  0.020 &  8.720  0.051 &  8.613  0.025 \\
\object{TYC 4770 1261 1}	& 12.232 0.160 & 11.498  0.106  & 10.721  0.048 & 10.458  0.046 & 10.299  0.042 \\
\object{TYC 4771 661 1}		& 12.386 0.201 & 11.899  0.178  & 10.617  0.026 & 10.298  0.026 & 10.225  0.024 \\
\object{TYC 4770 1432 1}	& 12.799 0.255 & 12.337  0.211  & 11.077  0.023 & 10.844  0.025 & 10.731  0.023 \\
\object{TYC 4770 1129 1}	& 12.866 0.240 & 11.827  0.146  & 11.232  0.021 & 11.034  0.024 & 10.979  0.023 \\
\object{TYC 4770 1018 1}	& 13.014 0.280 & 11.224  0.077  &  8.620  0.021 &  7.881  0.027 &  7.730  0.017 \\
\object{SO120532}$^{d}$	      	& 13.094 0.317 & 12.031  0.172  & 11.125  0.026 & 10.962  0.026 & 10.835  0.024 \\
\object{TYC 4771 1012 1}	& 13.009 0.302 & 11.263  0.090  &  8.998  0.019 &  8.365  0.061 &  8.205  0.026 \\
\object{TYC 4771 934 1}		& 13.325 0.387 & 12.870  0.306  &  8.948  0.032 &  8.374  0.057 &  8.207  0.026 \\
\object{TYC 4771 1468 1}	& 13.433 0.343 & 11.136  0.074  &  8.616  0.026 &  7.893  0.053 &  7.727  0.023 \\
\object{TYC 4770 924 1}		& 13.367 0.358 & 11.938  0.163  & 10.725  0.022 & 10.317  0.027 & 10.222  0.025 \\
            \noalign{\smallskip}
            \hline
            \noalign{\smallskip}
\object{$\sigma$ Ori D} 	&  6.375 0.012 &  6.557  0.014  &  7.116  0.029 &  7.219  0.027 &  7.260  0.021 \\
\object{$\sigma$ Ori C}$^{e}$ 	&  8.77  0.02  &  8.79   0.02   &  9.086  0.032 &  9.109  0.047 &  9.129  0.021 \\
\object{HD 294273} 		& 10.899 0.106 & 10.828  0.186  & 10.176  0.028 & 10.099  0.026 & 10.103  0.027 \\
\object{HD 294297} 		& 10.905 0.090 & 10.193  0.073  &  9.062  0.032 &  8.872  0.084 &  8.775  0.024 \\
\object{HD 294276} 		& 11.193 0.071 & 10.418  0.057  &  9.183  0.021 &  8.856  0.025 &  8.814  0.026 \\
            \noalign{\smallskip}
            \hline
         \end{tabular}
     $$ 
\begin{list}{}{}
\item[$^{a}$] Also \object{BD--02 1323C}. 
\item[$^{b}$] Also \object{2E 0537.1--0243}. 
\item[$^{c}$] Name from Wolk (1996).
Also \object{SO210457}. 
\item[$^{d}$] Name from Caballero (2006).
\item[$^{e}$] Johnson $B$- and $V$-band photometry from Greenstein \&
Wallerstein (1958)  
\end{list}
   \end{table*}
%

   \begin{table*}
      \caption[]{Astrometry, spectral type and radial velocity (when
      available) of all the stars.}  
         \label{estrellas_dos}
     $$ 
         \begin{tabular}{lccccccc}
            \hline
            \hline
            \noalign{\smallskip}
Name 				& TYC		& $\alpha$ 	& $\delta$ 	& $\mu_\alpha \cos{\delta}$& $\mu_\delta$ & Sp. 			& V$_r$ \\  
 				&		& (ICRS)	& (ICRS)	& $\pm \delta\mu_\alpha \cos{\delta}$ 	& $\pm \delta\mu_\delta$ & type & km\,s$^{-1}$ \\  
 				&		& 		& 		& [mas\,a$^{-1}$] & [mas\,a$^{-1}$] 	& 				& \\  
            \noalign{\smallskip}
            \hline
            \noalign{\smallskip}
$\sigma$ Ori A+B+IRS1$^{a}$&4771 1196 1 & 05 38 44.765& --02 36 00.25& +4.61$\pm$0.88 &--0.40$\pm$0.53 & O9.5V+...     & +29.1 		\\ 
$\sigma$ Ori E 		& 4771 1194 1   & 05 38 47.208& --02 35 40.52&   +2.2$\pm$1.1 & --1.4$\pm$1.2  & B2Vp	       & +29 		\\ 
HD 37699     		& 4771 1045 1   & 05 40 20.189& --02 26 08.22&   +0.3$\pm$1.2 &  +0.9$\pm$1.3  & B5Vn	       & +14.8$\pm$2.2 	\\ 
HD 294271    		& 4771 1193 1   & 05 38 36.549& --02 33 12.76&   +0.4$\pm$1.3 & --0.6$\pm$1.4  & B5V	       & 		\\ 
HD 37525 AB  		& 4771 1103 1   & 05 39 01.493& --02 38 56.36&   +2.1$\pm$1.3 & --0.4$\pm$1.3  & B5Vp	       & 		\\ 
HD 294272 A  		& 4771 1205 1   & 05 38 34.799& --02 34 15.78&  --3.3$\pm$1.6 &  +0.6$\pm$1.6  & B9.5III       & 		\\ 
HD 294272 B  		& 4771 1204 1   & 05 38 34.235& --02 34 16.08&   +5.4$\pm$1.8 & --3.4$\pm$1.7  & B8V	       & 		\\ 
HD 37333     		& 4771 956 1    & 05 37 40.476& --02 26 36.83&  --3.3$\pm$1.0 & --3.9$\pm$1.3  & A1Va	       & 		\\ 
HD 37564     		& 4771 1073 1   & 05 39 15.061& --02 31 37.62&   +2.2$\pm$1.2 &  +0.5$\pm$1.2  & A8V:	       & 		\\ 
V1147 Ori    		& 4771 685 1    & 05 39 46.196& --02 40 32.06&  --2.5$\pm$1.2 &  +4.0$\pm$1.3  & B9IIIp        & 		\\ 
HD 37686     		& 4771 909 1    & 05 40 13.090& --02 30 53.18&   +2.1$\pm$1.2 &  +2.1$\pm$1.2  & B9.5Vn        & 		\\ 
HD 37545     		& 4771 304 1    & 05 39 09.213& --02 56 34.73&   +1.9$\pm$1.1 & --1.1$\pm$1.1  & B9V	       & 		\\ 
HD 294275    		& 4771  39 1    & 05 37 31.872& --02 45 18.48&  --2.7$\pm$1.2 & --4.4$\pm$1.3  & A1V	       & 		\\ 
HD 294268		& 4771 802 1    & 05 38 14.121& --02 15 59.78&   +4.5$\pm$1.5 &  +0.8$\pm$1.5  & F5	       & +20 		\\ 
HD 294307    		& 4771 157 1    & 05 40 12.462& --02 52 57.59&   +0.1$\pm$1.4 &--16.8$\pm$1.4  & F8	       & 		\\ 
RX J0539.6--0242 AB	& 4771 543 1    & 05 39 36.541& --02 42 17.21&   +2.2$\pm$1.5 &  +2.8$\pm$1.6  & G5--K0        & +32.6$\pm$1.8 	\\ 
HD 294279    		& 4771 405 1    & 05 38 31.382& --02 55 03.15&  --0.4$\pm$1.4 &  +6.7$\pm$1.4  & F3--5         & 		\\ 
StHA 50  		& 4771 1087 1   & 05 38 34.449& --02 28 47.56&   +2.8$\pm$1.6 &  +3.2$\pm$1.6  & Be	       & 		\\ 
GSC 04771--00621	& 4771 621 1    & 05 37 41.793& --02 29 08.21&   +7.0$\pm$1.6 &--11.1$\pm$1.6  &	       & 		\\ 
HD 294277    		& 4771 463 1    & 05 37 57.342& --02 53 17.69&   +4.8$\pm$1.2 & --2.8$\pm$1.2  & K2	       & 		\\ 
HD 294278    		& 4771 127 1    & 05 38 38.762& --02 49 01.31&  +10.9$\pm$1.4 &  +2.6$\pm$1.4  & K2	       & 		\\ 
HD 294270    		& 4770 1336 1   & 05 37 18.818& --02 31 36.44&  +16.8$\pm$1.9 &--23.4$\pm$1.8  & G0	       & 		\\ 
HD 294301    		& 4771 556 1    & 05 40 21.125& --02 40 25.62&   +4.6$\pm$1.7 & --8.5$\pm$1.7  & F2V(n)        & 		\\ 
HD 294269    		& 4771 693 1    & 05 37 57.818& --02 26 33.65&  +43.5$\pm$1.8 &--16.1$\pm$1.8  & G0	       & +72.8 		\\ 
GSC 04771--00962	& 4771 962 1    & 05 37 44.915& --02 29 57.31&   +2.8$\pm$1.9 &  +2.7$\pm$2.1  &	       & 		\\ 
HD 294298    		& 4771 880 1    & 05 39 59.318& --02 22 54.35&  --0.4$\pm$1.8 &  +1.8$\pm$1.8  & G0:	       & +11.5 		\\ 
HD 294274    		& 4771 940 1    & 05 37 45.365& --02 44 12.53&  --8.1$\pm$3.2 &--19.4$\pm$3.5  & G0	       & 		\\ 
4771--0950		& 4771 950 1    & 05 38 06.496& --02 28 49.37&   +6.9$\pm$1.8 & --7.8$\pm$1.8  & F7	       & 		\\ 
HD 294280    		& 4771 385 1    & 05 38 28.491& --03 03 33.79&  --4.0$\pm$1.7 &  +4.9$\pm$1.8  & K5	       & 		\\ 
TYC 4771 720 1		& 4771 720 1    & 05 37 59.046& --02 41 00.47& --20.1$\pm$3.9 &--16.6$\pm$4.2  &	       & 		\\ 
2E 0535.4--0241		& 4771 921 1    & 05 37 54.405& --02 39 29.84&  --1.8$\pm$3.8 &  +0.9$\pm$4.0  &	       & 		\\ 
TYC 4770 1261 1		& 4770 1261 1   & 05 37 10.470& --02 30 07.19&  --4.3$\pm$3.8 & --8.5$\pm$4.1  &	       & 		\\ 
TYC 4771 661 1		& 4771 661 1    & 05 39 20.446& --02 27 51.44&  +16.6$\pm$4.3 &--20.1$\pm$4.7  &	       & 		\\ 
TYC 4770 1432 1		& 4770 1432 1   & 05 37 09.689& --02 39 59.87& --30.9$\pm$4.5 &  +0.1$\pm$5.0  &	       & 		\\ 
TYC 4770 1129 1		& 4770 1129 1   & 05 37 08.495& --02 22 06.34&  --5.8$\pm$4.2 &  +0.4$\pm$4.7  &	       & 		\\ 
TYC 4770 1018 1		& 4770 1018 1   & 05 36 57.142& --02 25 40.00&  --4.3$\pm$4.2 & --2.3$\pm$4.6  &	       & 		\\ 
SO120532	      	& 4771 873 1    & 05 38 14.438& --02 19 58.64&  --3.0$\pm$4.3 & --5.5$\pm$4.7  & F7--9         & 		\\ 
TYC 4771 1012 1		& 4771 1012 1   & 05 39 13.944& --02 10 49.27&  --0.9$\pm$4.2 & --9.7$\pm$4.6  &	       & 		\\ 
TYC 4771 934 1		& 4771 934 1    & 05 39 43.058& --02 28 45.59&  --7.5$\pm$4.6 & --2.7$\pm$4.9  &	       & 		\\ 
TYC 4771 1468 1		& 4771 1468 1   & 05 37 29.879& --02 43 45.83&  --7.0$\pm$4.7 & --6.5$\pm$4.9  &	       & 		\\ 
TYC 4770 924 1		& 4770 924 1    & 05 37 19.240& --02 28 46.22&   +1.8$\pm$4.7 &--43.4$\pm$5.1  &	       & 		\\ 
            \noalign{\smallskip}
            \hline
            \noalign{\smallskip}
$\sigma$ Ori D$^{b}$	& 4771 1195 1   & 05 38 45.602& --02 35 58.84&   +4.6$\pm$0.9 & --0.4$\pm$0.5  & B2V 	       & +33.1$\pm$3.1 	\\ 
$\sigma$ Ori C$^{c}$ 	& -- 		& 05 38 44.12 & --02 36 06.3 &      0$\pm$0   &     0$\pm$0    & A2V           & 		\\ 
HD 294273$^{b}$ 	& 4771 1023 1   & 05 38 27.524& --02 43 32.60&  --0.6$\pm$2.8 &  +4.6$\pm$2.5  & A3            & 		\\ 
HD 294297$^{b}$ 	& 4771 1081 1   & 05 40 27.531& --02 25 42.90&  +22.7$\pm$2.6 &--20.5$\pm$1.7  & F6--8 	       & +25 		\\ 
HD 294276$^{b}$ 	& 4770 1489 1   & 05 37 20.671& --02 49 32.55&   +8.8$\pm$1.7 &--61.7$\pm$3.6  & G0            & 		\\ 
            \noalign{\smallskip}
            \hline
         \end{tabular}
     $$ 
\begin{list}{}{}
\item[$^{a}$] Proper motions from Perryman et al. (1997). 
The spectral types of the components of the triple system are O9.5V, B0.5V and K:.
\item[$^{b}$] Proper motions from H{\o}g et al. (1998). 
\item[$^{c}$] Coordinates from the 2MASS catalogue and proper motion from the
NOMAD1 catalogue.  
\end{list}
   \end{table*}